\documentclass[twocolumn,trackchanges]{aastex702}
\usepackage[T1]{fontenc}
\usepackage[utf8]{inputenc}
\usepackage{multirow}
\usepackage{graphicx}
\usepackage{amsmath}
\usepackage[normalem]{ulem}
\usepackage{mathrsfs}
\usepackage{hyperref}
\usepackage{threeparttable}
\usepackage{booktabs}
\usepackage{float}
\usepackage{array}
\usepackage{bm}
\usepackage{soul}
\usepackage{xspace}
\usepackage[colorinlistoftodos]{todonotes}
\usepackage{booktabs}
\usepackage{siunitx}
\usepackage{rotating}
\usepackage{graphicx}

\begin{document}

\title{Reassessment of Planetary Parameters and Statistical Validation of 187 Kepler Habitable-Zone and Near Habitable Confirmed and Candidate Planets}

\author[orcid=0000-0002-3696-2127,sname='North America']{Md Abdullah Al Zaman}
\affiliation{Department of Physics and Astronomy, University of Kansas, Lawrence, KS, USA}
\email[show]{abdullah.alzaman@ku.edu}

\author[sname='North America']{Ian J. M. Crossfield}
\affiliation{Department of Physics and Astronomy, University of Kansas, Lawrence, KS, USA}
\email[hide]{ianc@ku.edu}  

\author[orcid=0000-0002-6191-459X,sname='North America']{Ekrem Murat Esmer}
\affiliation{Department of Physics and Astronomy, University of Kansas, Lawrence, KS, USA}
\email[hide]{ianc@ku.edu}




\author[orcid=0009-0000-2075-1109,sname='North America']{Jason Pero}
\affiliation{Department of Physics and Astronomy, University of Kansas, Lawrence, KS, USA}
\email[hide]{ianc@ku.edu}

\begin{abstract}

{Confirmed and candidate planets in the Habitable or near-Habitable Zones (HZ) orbiting Sun-like stars are crucial not only for the search for life beyond the Solar System, but also for understanding the architecture and formation of other solar systems and for conducting demographic studies.  Current analyses of such planetary systems are limited both by the small sample size revealed by {\em Kepler} as well as by uncertainties in system parameters.  Although a larger sample of HZ and near-HZ planets must await new observations from wide-field, space-based surveys, follow-up analyses and observations with existing facilities can improve the precision on stellar and planetary parameters, enable more accurate planetary classification, and a more reliable assessment of whether candidate systems are true planets or merely astrophysical false positives.
Here, we present a homogeneous transit light curve analysis of 187  HZ and near-HZ systems discovered by {\em Kepler}. Our analysis incorporates the latest Gaia DR3-derived stellar parameters, and we also present a statistical validation effort for all targets.  The result is a uniform set of planetary radii,  instellation fluxes, and orbital ephemerides suitable for prioritizing and scheduling follow-up observations to improve our understanding of planets in this regime.  We show that longer-period ($P\gtrsim300$\,d) systems would particularly benefit from transit follow-up: just a single transit from facilities such as the 6\,m {\em James Webb} or the planned 3\,m {\em Lazuli} space telescopes could provide $\sim$2$\times$ tighter precision on planet radii than four years of {\em Kepler} photometry, thereby dramatically shrinking relevant parameter uncertainties and allowing more confident planet/false-positive assessments. 

}


\vspace{10pt}
\textbf{\large *****Accepted to AJ on 13th November}

\end{abstract}

\section{Introduction} 
The Habitable Zone (HZ) conventionally refers to the circumstellar region where a rocky planet with an Earth-like atmosphere is expected to have surface temperatures that allow liquid water to exist \citep{kasting1993,kopparapu2013habitable}. This region is typically defined using key orbital and stellar parameters, including the semi-major axis, orbital period, stellar spectral type, and the instellation (or insolation) flux ($S_\oplus$) received by the planet \citep{petigura2013prevalence}. Conservative estimates suggest that interior to the HZ, photolysis and hydrogen escape lead to permanent water loss, whereas beyond the HZ, the formation of $CO_2$ clouds cools the planet’s surface by increasing the albedo and reducing the convective lapse rate. 

Stellar luminosity is one of the key aspects for defining the HZ. Less luminous stars possess HZs that are smaller in size and the planets orbiting there occur at closer orbital distances \citep{kasting1993}. The boundaries of the HZ are expected to evolve outwards as the star ages and becomes more luminous. The so-called optimistic habitable zone is calculated under the assumption that Venus once possessed liquid water on its surface early in its history, but later lost it \citep{turbet2021day}, while the outer edge is determined based on the evidence that early Mars sustained surface liquid water \citep{Amri,kopparapu2013habitable,kopparapu2014habitable}. However, not all planets in the habitable zone (whether conservative or optimistic) are necessarily habitable, and not all habitable planets must reside within the HZ of their host stars. Several other factors — such as the presence of a planetary magnetic field \citep{airapetian2017hospitable,tilley2019modeling}, the planet’s water content \citep{cowan2014water,kurosaki2014impact}, the influence of plate tectonics \citep{noack2014can}, the core mass fraction \citep{noack2017water}, and the level of radiation \citep{al2023effectiveness} — all contribute to determining a planet’s habitability. In addition, rocky planets within the habitable zone generally exhibit low eccentricities \citep{hill2023catalog}, although planets with radii between 1.5–2.0 $R_\oplus$ may display elevated eccentricities \citep{gilbert2025planets}.

Detecting earth-sized/mass planets orbiting Sun-like stars with the potential to harbor life has been a longstanding goal of the exoplanet community. The Kepler space telescope, launched
in 2009, with its unprecedented photometric precision, identified thousands of exoplanets by monitoring the brightness of over 150,000 stars over four years \citep{borucki2010Kepler,batalha2014exploring,koch2010Kepler}, revolutionizing our understanding of planetary systems and the diverse population of exoplanets. One of the high-priority objectives of the Kepler mission was to identify small, rocky exoplanets ($\lesssim 1.8\, R_\oplus$)  that receive stellar irradiation levels conducive to sustaining surface liquid water. High stellar variability, noise, and instrumental systematics \citep{gilliland2011Kepler} caused many of those planets to fall below the detection threshold. Some of the detections also suffer from poor reliability \citep{2018ApJS..235...38T,bryson2020occurrence}.

Studying HZ planets is essential for probing planetary system architectures and for constraining exoplanet occurrence rates, particularly Eta-Earth \citep[$\eta_\oplus$;][]{lunine2008worlds}, defined as the occurrence rate of rocky planets orbiting Sun-like stars within their HZs. Following the Kepler mission, studies of $\eta_\oplus$ have reported values from less than 0.1 to 1 with
large uncertainties \citep{bryson2025estimating,fernandes2025we,bergsten2022demographics,10.1088/2514-3433/abfa8fch2,bryson2020occurrence,bryson2020probabilistic,kunimoto2020searching,hsu2019occurrence,mulders2018exoplanet,burke2015terrestrial,silburt2015statistical,foreman2014exoplanet,petigura2013prevalence}. The large uncertainties in $\eta_\oplus$ estimation stem from several factors. These include limited sample sizes and insufficient numbers of observed transits, as well as differences in the regimes of planetary properties such as radius, orbital period, and instellation flux \citep{10.1088/2514-3433/abfa8fch2,bryson2020occurrence,kunimoto2020searching}. Additionally, variations in stellar properties and methodological differences in algorithms, treatments, and extrapolations contribute to the uncertainties. Furthermore, we have not yet identified a true Earth analog orbiting a Sun-like star \citep{2026MNRAS.547ag028B}.



Refining planetary parameters is essential for obtaining robust and unbiased estimates of $\eta_\oplus$. Accurate parameter determination reduces the risk of planet misclassifications, particularly in regimes where detection uncertainties or stellar parameter errors can blur the distinction between terrestrial and sub-Neptune-sized planets. As $\eta_\oplus$ is derived from population statistics sensitive to both planetary radius and instellation flux, even modest improvements in stellar and planetary characterization can significantly alter inferred occurrence rates within the HZ \citep{kunimoto2020searching}. Improved planetary parameters also allow better calibration of planet formation and evolution models, helping constrain theories on core accretion, atmospheric loss, and migration mechanisms, which ensures that simulated populations align more closely with observed distributions. This refinement enables more coherent comparison between datasets employing different techniques such as transit, radial velocity (RV), direct imaging, and microlensing and mitigates systematic offsets between catalogs. A refined parameter space further allows a more accurate and physically consistent delineation of planets within the HZ, thereby improving the reliability of habitability assessments \citep{stalport2022general}. In short, parameter refinement forms the backbone of the empirical framework linking planetary demographics to the search for life beyond the Solar System.

\begin{figure*}[t!] 
    \centering
    \includegraphics[width=0.9\linewidth]{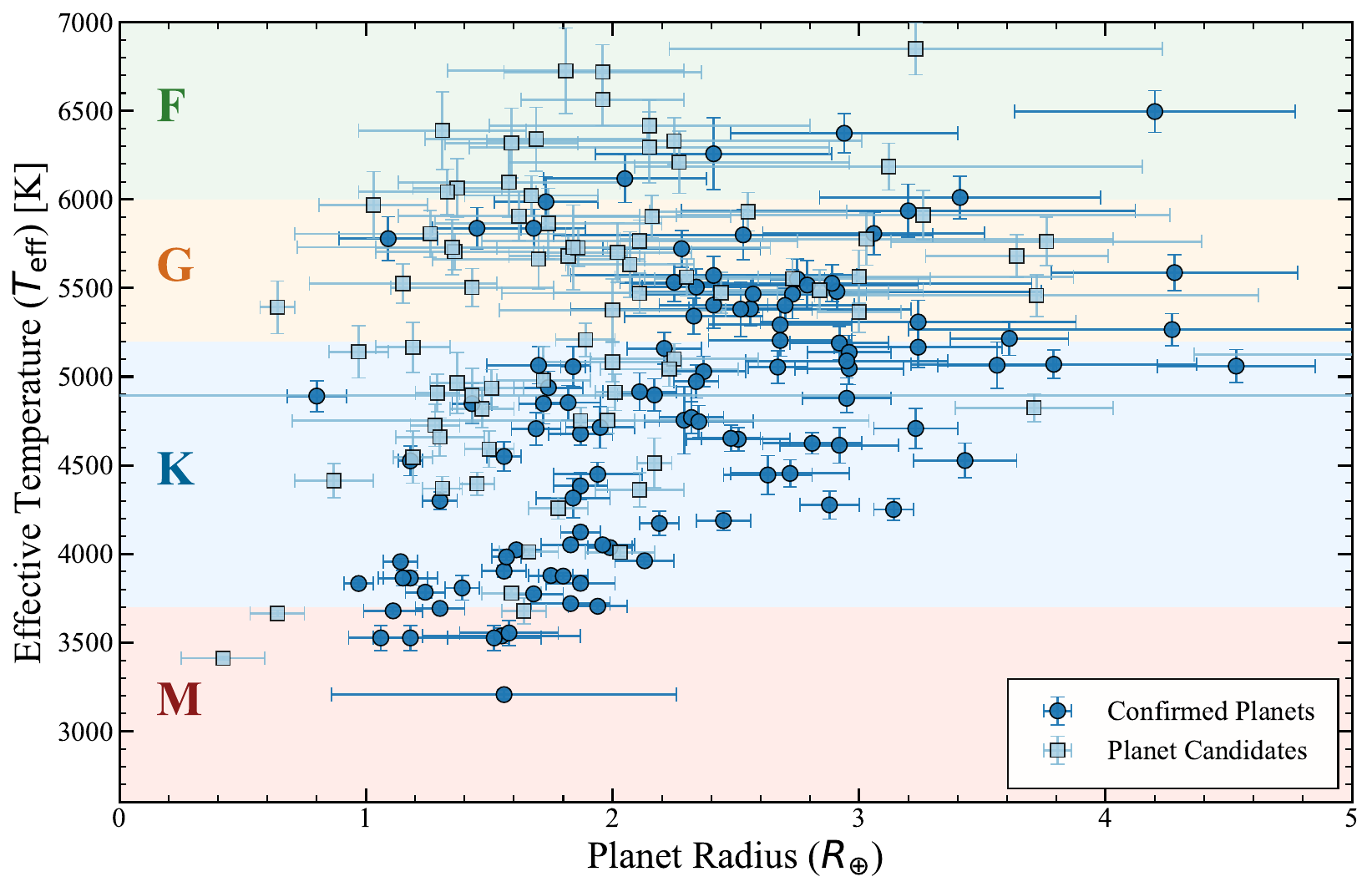}
    \caption{Distribution of planets in our sample in terms of stellar effective temperature and planetary radius \citep{thompson2018planetary}. The shaded regions indicate different stellar spectral types (F, G, K, and M), and the symbols distinguish confirmed planets (circles) from planet candidates (squares). The definition of confirmed planets and planet candidates is described in Section ~\ref{sec:my_sample}}
    \label{fig:planet_radius_vs_teff}
\end{figure*}

{The primary objectives of this paper are to present (i) a homogeneous reanalysis of all known Kepler habitable-zone (HZ) and near-HZ planets, (ii) the incorporation of updated stellar parameters for all systems, and (iii) an independent derivation of false-positive probabilities for all systems. 
The revised catalog is intended to improve our understanding of the distribution of temperate terrestrial worlds and provide a uniform set of system parameters to prioritize future follow-up and analysis efforts. Section~\ref{sec:my_sample} presents our sample. Section~\ref{sec:analytical_framework} discusses our analytical framework, highlighting the differences in our stellar parameters, detrending techniques relative to Kepler DR25, and the statistical validation framework. Section~\ref{sec:results} describes the results, discussion and outlines directions for future work, and Section~\ref{sec:conclusion} provides the conclusions.

\section{Sample}
\label{sec:my_sample}
In this study, we selected a sample of 187 planets located within or near the habitable zone (HZ). This sample was constructed as the union of the habitable-zone, terrestrial-sized candidates identified by Kepler Q1-17 Data Release 25 \citep[DR25;][]{2018ApJS..235...38T}; the rocky, habitable-zone planet population used in the occurrence-rate analysis of \citep{2021AJ....161...36B}; the Habitable table Zone Exoplanets catalog \citep{hill2023catalog}; planets around M/K type stars with Earth-like sizes and Instellation flux \citep{kipping2025near} and the superhabitable candidate worlds described by \cite{schulze2020search}. Of this sample, 108 objects are confirmed planets, having measured masses and sufficient follow-up observations for validation; the remaining are classified as planet candidates under the Exoplanet Archive disposition, which is derived from the Disposition Using Kepler Data. The NASA Exoplanet Archive's pipeline-based disposition is assigned through a series of false-positive tests applied to Kepler pixel-level and light-curve data, as described in \cite{batalha2013planetary}, and implemented in later releases using the Kepler Q1-17 DR25 Robovetter \citep{coughlin2020robovetter} for the automatic vetting of Threshold Crossing Events (TCEs). The reliability values of the confirmed planets and candidates range from 0.01 to 1.00 \citep{2021AJ....161...36B}. Of these, 178 can be found in the Kepler Q1–17 DR25 catalog \citep{2018ApJS..235...38T}, while the remainder come from the Kepler Q1-17 Data Release 24 (DR24) \citep{coughlin2016planetary} and Kepler Q1–16 \citep{mullally2015planetary} lists. The Habitable Zone Exoplanets catalog and the superhabitable-world sample (planets with a high likelihood of harboring life) \citep{hill2023catalog,schulze2020search} include several relatively bigger ($>7_\oplus$) planets (KOI-375.01, KOI-433.02, KOI-3762.01, KOI-3791.01, KOI-5084.01, KOI-5416.01. Several planets in these catalogues are flagged as false positives (KOI-5009.01, KOI-5789.01, KOI-5870.01, KOI-5878.01, KOI-5879.01, KOI-5896.01, KOI-5715.01, KOI-5389.01) in the Kepler Q1–Q16 catalogue. Although they were discarded in the DR25 catalogue, subsequent analysis reintroduced them in the Q1–Q17 DR25 Supplemental Table (as candidates). However, we did not get the updated planetary parameter values from the Q1–Q17 DR25 Supplemental Table; therefore, we excluded these {Kepler Objects of Interest (KOIs)}} from our analysis.

The planets exhibit orbital periods ranging from 4 to 624 days, with an average of 185 days. Their radii span from 0.4 $R_\oplus$ to 6.13 $R_\oplus$, with an average of approximately 2.15 $R_\oplus$. They orbit main-sequence stars of spectral types F, G, K, and M, with stellar effective temperatures between 3200 K and 6800 K. Figure \ref{fig:planet_radius_vs_teff} illustrates the distribution of planets in our sample with respect to their effective temperatures and radii. The relatively long orbital periods and large semi-major axes of most of the planets in the sample imply that these are unlikely to be tidally locked, potentially maintaining Earth-like rotational and orbital dynamics. Planets receiving instellation flux near 1 $S_\oplus$ may sustain atmospheres with greenhouse gases capable of maintaining stable surface conditions, provided they possess sufficient atmospheric pressure and composition stability, while those below 0.5 $S_\oplus$ might require elevated CO$_2$ concentrations or thick atmospheres to avoid global freezing \citep{forget1997warming,kopparapu2013habitable}. Since some of the host stars range from late K to M dwarf types, the corresponding habitable zones are closer to the star, making tidal locking and stellar activity \citep{mota2025habitable} key factors in habitability and atmospheric retention. Planets with Instellation flux over 1.80 $S_\oplus$ may possess a dense secondary atmosphere, resulting in Venus-like or even hotter surfaces. Those planets lie within the region where worlds transition from potentially habitable to inhabitable states due to the runaway greenhouse mechanism. Because of this aspect, some of these planets (Such as KOI-250.04, KOI-581.02, KOI-2931.01, KOI-8159.02, KOI-2290.01, etc.) are relevant in  \cite{bryson2020occurrence} for the occurrence-rate analysis.

\setlength{\tabcolsep}{4pt}

\begin{deluxetable*}{rll}
\tabletypesize{\scriptsize}
\tablecaption{Function, Prior Bounds, and Convergence Metrics for MCMC Transit Modeling}\label{tab:mcmcsettings}
\tablewidth{\linewidth}
\label{tab:functions_and_bounds}
\tablehead{
  \colhead{Parameter} & 
  \colhead{Prior Type} & 
  \colhead{Bound}
}
\startdata
$R_p/R_*$        & Uniform Prior & $(R_{p}/R_{*})_{guess} \pm 50\%$ \\
Period ($P$)     & Uniform Prior & $P_{\rm guess} \pm 1\%$ \\
$T_0$            & Uniform Prior & $T_{0,\rm guess}\pm 0.1days$\\
$a/R_*$          & Uniform Prior & $(a/R_*)_{\rm guess} \pm 40\%$ \\
Inclination ($i$) & Geometric Prior in $\sin i$   & $87^\circ$--$90^\circ$ \\
$u_1$, $u_2$     & Gaussian Priors & Standard form \\
\hline
\multicolumn{3}{l}{\textbf{Convergence Check Metrics:}} \\
\multicolumn{3}{l}{Gelman-Rubin ($\hat{R}$) statistic: $\hat{R} < 1.10$} \\
\multicolumn{3}{l}{Autocorrelation Length $N/\tau$>50}\\
\multicolumn{3}{l}{Acceptance fraction: $0.20 - 0.45\%$} \\
\multicolumn{3}{l}{Convergence and trace plot monitoring} \\
\enddata
\tablecomments{
The orbital eccentricity ($e$) is assumed to be zero and treated as a fixed parameter, corresponding to a perfectly circular orbit. The log-likelihood function uses the observed flux uncertainties and uses the standard $\chi^2$ form. No photometric jitter parameter was included in the model. The prior distributions used in the Bayesian inference were informed by values reported in the DR25 and DR Q1–Q16 catalogs. 
}
\end{deluxetable*}

\section{Analytical Framework}
\label{sec:analytical_framework}
\subsection{Stellar Parameters}
\label{sec:stellar}
Precisely constraining the planetary parameters requires precise stellar parameters \citep{mortier2013new}. Different methods used to estimate stellar parameters can produce systematically different results \citep{berger2023gaia}, leading to variations in the inferred planetary properties. Precise stellar parameters such as effective temperature ($T_{eff}$), metallicity, and surface gravity ($log g$) are essential for several aspects of exoplanet analysis. They help define the location of the HZ \citep{bryson2020occurrence,bryson2025estimating}, estimate limb darkening coefficients, interpret transit observations, and understand stellar activity. The planetary mass and radius depend directly on the host star’s mass and radius. In addition, stellar properties are linked to protoplanetary disk characteristics, planet formation pathways, and planet formation efficiency, and they provide important constraints on planetary evolution and system age. They also serve as key inputs for modeling planetary atmospheres, including atmospheric chemistry, temperature structure, and potential atmospheric loss.

Kepler DR25 utilized stellar parameters from \cite{mathur2017revised}, which relied on a heterogeneous approach combining spectroscopy, photometry, asteroseismology, and stellar models available at that time. The stellar parameters were drawn from a variety of sources, including Kepler’s own Kepler Input Catalog (KIC), the Large Sky Area Multi-Object Fiber Spectroscopic Telescope (LAMOST), the Apache Point Observatory for Galactic Evolution Experiment (APOGEE), the Kepler Community Follow-up Program (CFOP), and several published papers. For 14,535 stars, they also adopted surface gravities estimated from the detection of granulation in Kepler light curves using the Flicker method \citep{bastien2016granulation}. This heterogeneous data set could introduce systematic biases into key stellar parameters such as $T_\mathrm{eff}$ and stellar radius ($R_*$), which in turn would propagate into derived planetary radii and instellation fluxes. 

We therefore draw our stellar parameters from the updated Gaia-Kepler-TESS Host Stellar Properties Catalog \citep{berger2023gaia}, which combines Gaia DR3 photometry, parallaxes, and spectroscopic metallicities. These stellar parameters were derived with a uniform isochrone-fitting pipeline to reduce systematic offsets. In comparison with studies based on fundamental and precise constraints such as interferometry and asteroseismology, good agreement is found between the $T_{\rm eff}$, radii, and ages reported in this catalog and those in the literature \citep{berger2023gaia}. Since Gaia parallaxes and photometry directly constrain stellar luminosity and radius, the derived instellation flux ($S_\oplus$) and habitable-zone boundaries become significantly more reliable, enabling more precise placement of HZ and near-HZ confirmed and candidate planets. 
However, we note that ten of our systems (KOI-2529.02, KOI-2626.01, KOI-1422.02, KOI-1422.04, KOI-1422.05, KOI-8000.01, KOI-3010.01, KOI-2862.01, KOI-3255.01, and KOI-7235.01) were not included in \cite{berger2023gaia}; for these, we adopt their stellar parameters from the Kepler Q1–17 DR25 catalog \citep{2018ApJS..235...38T}. 

Using these stellar parameters, we estimated the quadratic limb-darkening coefficients with Limb Darkening Toolkit \citep[\texttt{LDTk};][]{ldtk}, employing the PHOENIX model-spectra grid \citep{husser2013new} with the visible spectrum (VIS) dataset within the Kepler bandpass. In this process, for planets KOI-463.01, KOI-2290.01, KOI-2650.01, KOI-817.01, \texttt{LDTk} did not yield valid coefficients (0.1 with no uncertainty values); in those cases, we adopted values computed with Exoplanet Characterization Toolkit (\texttt{ExoCTK}) \citep{exoctk} with the Kepler bandpass for one channel (stellar intensity over the entire bandpass at once). In this work, for both \texttt{LDTk} and \texttt{ExoCTK}, we adopt limb-darkening parameter uncertainties derived directly from the covariance matrix of the coefficients obtained from fitting the analytical laws to the model atmosphere intensity profiles.

\subsection{Detrending and Bayesian Inference}
Our analysis employed a custom code called Synthetic Transit Lightcurve HarvestER for Exoplanet Science Cases (\texttt{STERNA}) (Zaman et al. 2026, in preparation), which primarily serves as a simulated light-curve generator for observation planning. It was originally developed to simulate transit scenarios for the upcoming \textit{Lazuli Space Observatory} \citep{roy2026lazuli}, which employs an off-axis Three-Mirror Anastigmat (TMA) optical design, but it can also be adapted for other telescope configurations. One of its modules also enables the analysis of real observations, such as those obtained from \textit{Kepler} or \textit{TESS}. We used the \texttt{LightKurve} \citep{lightkurve} pipeline to download the Kepler long-cadence (30-minute) Pre-search Data Conditioning and Simple Aperture Photometry (PDCSAP) data. 

\begin{figure}[tb!]
    \includegraphics[width=\linewidth]{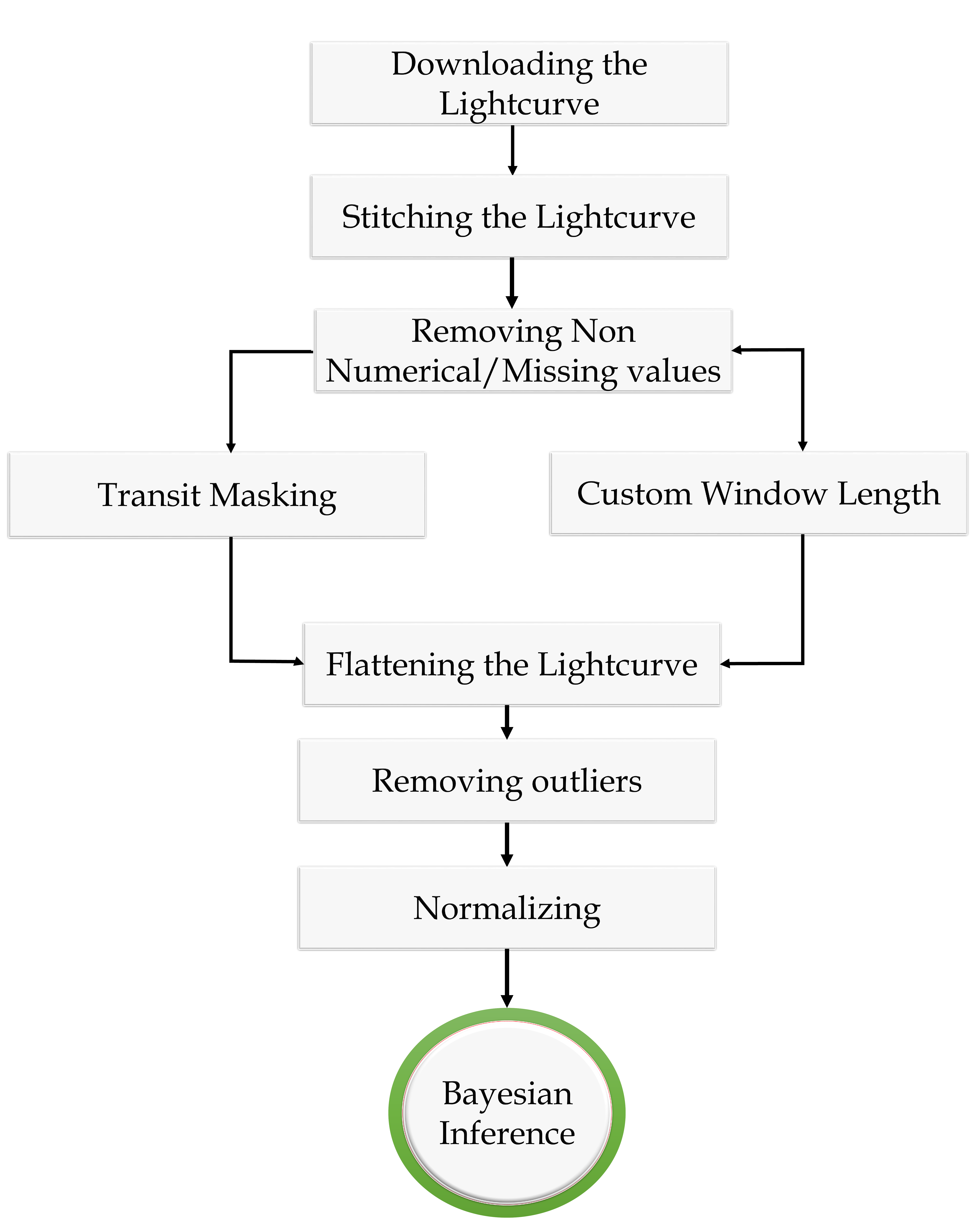}
    \caption{Detrending workflow applied to the light curve used in this study.}
    \label{fig:detrending_workflow}
\end{figure}

The detrending procedure for this work is summarized in Figure \ref{fig:detrending_workflow}. In Kepler DR25, \cite{thompson2018planetary} used the detrending procedure mentioned in \cite{rowe2014validation}, which starts with fitting the light curve (PDCSAP) with a cubic polynomial avoiding the transit window. To derive planetary parameters, they used a multi‑planet photometric model (assuming non‑interacting circular orbits and the quadratic limb‑darkening transit model) described by \cite{mandel2002analytic} to fit the detrended light curve. In our work, after downloading the data, stitching is done by the default stitch function available in the \texttt{LightKurve} package. We masked the transit (all the data within one transit duration) before detrending the light curves to prevent the detrending procedure from distorting the transit signal. To flatten each light curve, we applied a Savitzky–Golay filter \citep{savitzky}. The window length was determined by converting the transit duration of the confirmed/planet candidates into the corresponding number of 30-minute cadence points and setting the Savitzky–Golay filter window length to 15 times this value, rounded to the nearest odd integer as required by the filter. 
It was not arbitrarily chosen; we tested several window lengths, ranging from 10 to 18 times the transit duration, and found that a window length of 15 times the transit duration provided a suitable balance. It enables the analysis of all targets within a common framework, ensuring a homogeneous and consistent analysis. We use the default value (5$\sigma$) of \texttt{LightKurve} to remove the outlier. Also, we analyzed each planet in any multi-planetary system separately rather than fitting all planets simultaneously. For example, for the KOI-1422 system, we first analyzed KOI-1422.02 using the window length times its transit duration. After completing the analysis of KOI-1422.02, we proceeded to KOI-1422.04 and then KOI-1422.05, using the corresponding transit duration for each planet. 


The detrended light curve is passed into a Bayesian inference framework. The process begins by modeling the transit light curve using \texttt{BATMAN} \citep{batman}, which computes the analytic quadratic limb-darkened transit profile of a planet orbiting a limb-darkened star. 

To ensure robust estimates, carefully constructed prior probabilities are applied—including Gaussian priors on the limb-darkening coefficients, adaptive uniform bounds on geometric and timing parameters, and a geometric prior on the orbital inclination, proportional to sine. The details of the process are described in Table ~\ref{tab:functions_and_bounds}. The adopted prior boundaries were not arbitrarily selected; instead, they were empirically motivated and designed to incorporate the available observational constraints. For each system, the prior boundaries were centered around the reported values from the DR25 and Q1-Q16 catalogs, with widths chosen to reflect the associated measurement uncertainties while allowing sufficient parameter freedom during the exploration. Across the full sample of planets and planet candidates, these prior boundaries consistently resulted in well-converged chains and consistent posterior estimates, demonstrating their reliability within our Bayesian framework. The likelihood function quantifies the fit between the observed photometric data and the transit model, assuming Gaussian uncertainties. Posterior exploration is performed via Markov Chain Monte Carlo (MCMC) sampling with the  \citep[\texttt{emcee};][]{emcee} ensemble sampler, and automated convergence checks are based on the Gelman–Rubin statistic ($<1.1$) \citep{gelman} and the acceptance fraction ($20-45\%$) \citep{acceptance}. Posterior exploration was performed using 32 walkers and 50,000 steps per walker, discarding the first $20\% $ of each chain as burn-in. We use \texttt{ArviZ} \citep{kumar2019arviz} to extract the inference data for all the parameters and variables that we fit with \texttt{emcee}. From the different chains of samples, \texttt{ArviZ} calculates the R-hat ($\hat{R}$ ). If any parameter had $\hat{R} > 1.10 $ after the initial run, the code performed up to five additional runs of 10,000 steps each, centering the walkers on the median of the previous posterior and updating the burn-in fraction to $20\%  $ of the cumulative chain length. Final parameter values and uncertainties are reported as the median and the 16th/84th percentiles of the converged posterior samples after the last successful run.


We also investigated whether the chain lengths were sufficiently long to satisfy the $N>50\tau$ \citep{goodman2010ensemble}, where N is the total number of post-burn-in steps per walker and \(\tau \) is the integrated autocorrelation length. On average, the $N/\tau$ value was 207.2 with a median value of 240.87. In the majority of the cases, the $\hat{R}$ values for all the parameters were well below 1.05. We also ran the MCMC sampling on a subset of 20 randomly selected confirmed and candidate planets from our full sample with 
\(\hat{R} < 1.01\)\ \citep{vehtari2021rank}. This was done to check for substantial changes in the parameter values or their precision; however, no significant differences were observed.


From the results of the Bayesian inference analysis, we derived the probabilistic planetary mass ($M_p$) using  \texttt{Spright} \citep{parviainen2024spright} together with the estimated planetary radius ($R_p$).  We derived the semi-major axis ($a$) and instellation flux $S_\oplus$ of the planets in our sample using Equations (1) and (2) below:
\begin{equation}
a = \left( \frac{G M_* P^2}{4 \pi^2} \right)^{1/3}
\tag{1}
\end{equation}
where $G$, $M_*$ and $P$ are the gravitational constant, stellar mass and orbital period respectively, and
\vspace{1pt}
\begin{equation}
S = \frac{L_*}{4 \pi a^2}
\tag{2}
\end{equation}
where $L_*$ and $a$ are the stellar luminosity and semi-major axis, respectively.
\vspace{1pt}

However, for several planets in this sample, we observed significant stellar variability, resulting in noticeably noisier light curves. Although our initial fits produced results, we refined the analysis for these systems to improve accuracy. Specifically, we extracted segments spanning approximately five times the transit duration, detrended each segment using a second-degree polynomial, and then stitched the corrected portions back together to produce a more stable light curve (Figure ~\ref{fig:fitting}). We then applied the same fitting procedure described in Figure \ref{fig:detrending_workflow}. A total of fifteen planets (KOI-8077.01, KOI-8047.01, KOI-2124.01, KOI-8156.01, KOI-4014.01, KOI-1608.03, KOI-7923.01, KOI-2172.02, KOI-5433.01, KOI-4015.01, KOI-581.02, KOI-5581.01, KOI-1989.01, KOI-8174.01, KOI-2686.01) in this sample were processed using this method. In these cases, we found that the average acceptance fraction was $\sim35$ percent, and the $\hat{R}$ values were below 1.03.

\begin{figure*}[t!]
    \includegraphics[width=1\linewidth]{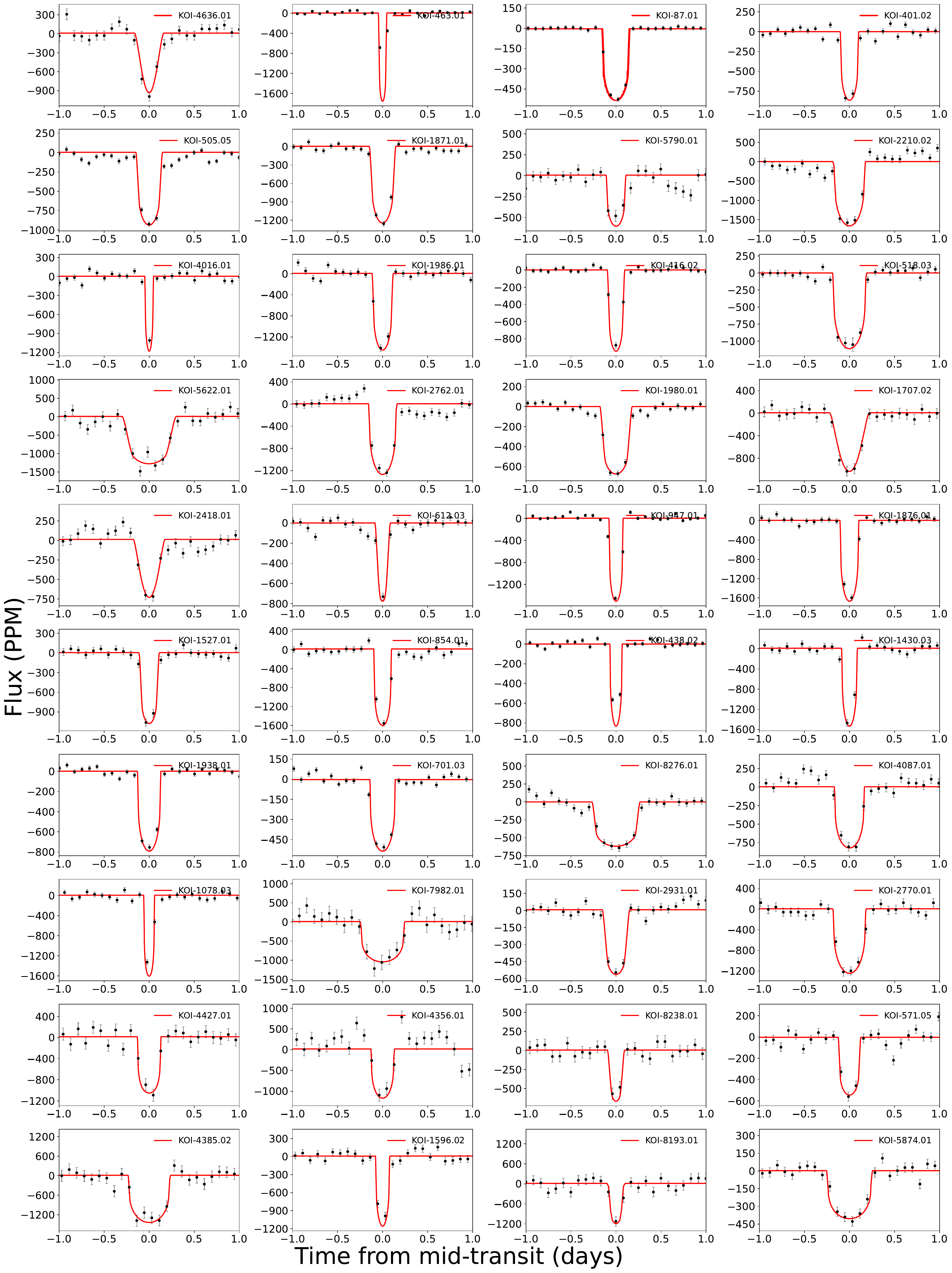}
    \caption{Transit fits for the first 40 planets/planet candidates from the sample of 187. The data are binned to 0.084 days ($\sim$2\, hours). Please see the appendix for all fits.}
    \label{fig:40_transit_fits}
\end{figure*}

\begin{figure*}[t!]
    \includegraphics[width=1\linewidth]{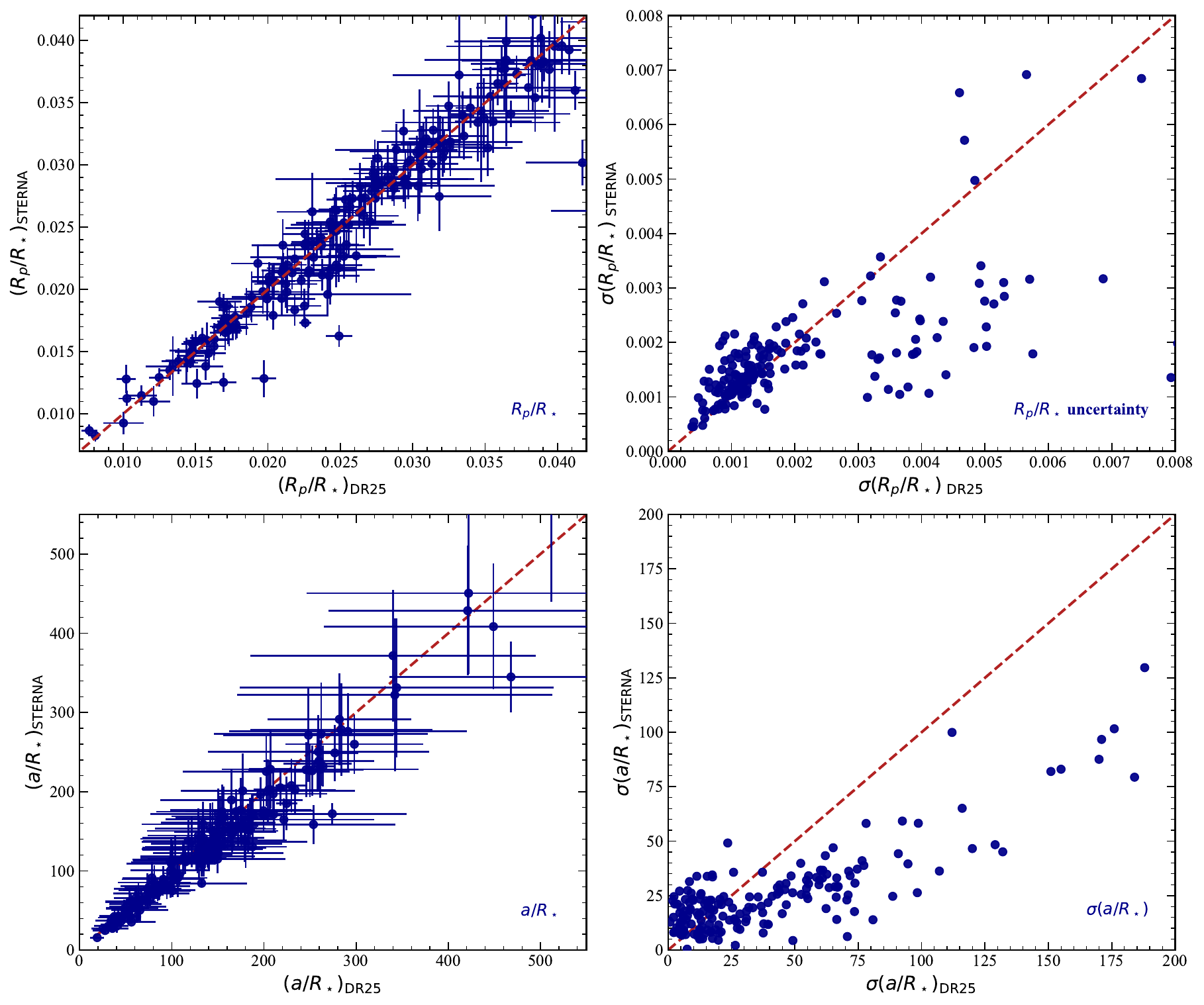}
    \caption{Our values are compared with those reported in the Kepler Q1--Q17 DR25 catalog. Extreme outliers (KOI-4636.01, KOI-1707.02, KOI-2704.01) are omitted for clarity. The two panels on the left compare the planet-to-star radius ratio ($R_p/R_*$) and the scaled semi-major axis ($a/R_*$), derived in this work, with the corresponding values from Kepler DR25. The two panels on the right show the improvement in parameter constraints by the STERNA fits relative to the DR25 catalog for both $R_p/R_*$ and $a/R_*$.}
    \label{fig:comparison}
\end{figure*}

\begin{figure}[tb!]
    \includegraphics[width=1\linewidth]{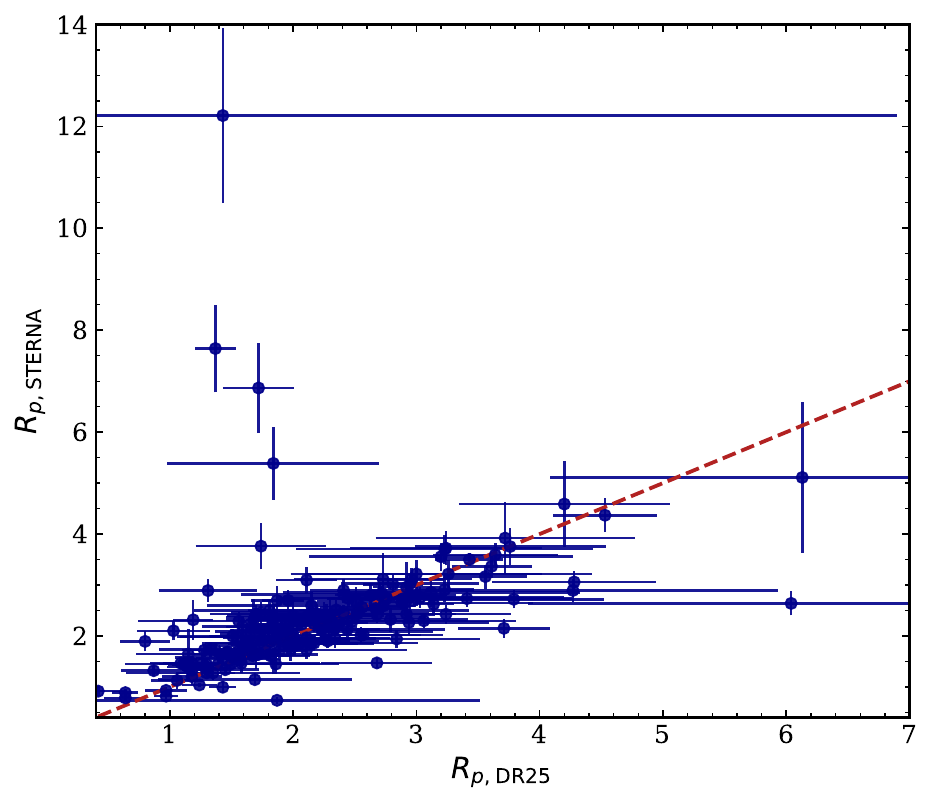}
    \caption{Our values vs Kepler Q1-17 DR25 data comparison for radius of the confirmed/candidate planet ($R_{\oplus}$). {The largest  deviations from DR25 values are primarily due to the updated stellar parameters (in this case the radius of the star $R_{\odot}$); see Sec.~\ref{sec:revised}.}}
    \label{fig:comparison_radius}
\end{figure}

\begin{figure*}[t!]
    \includegraphics[width=0.95\linewidth]{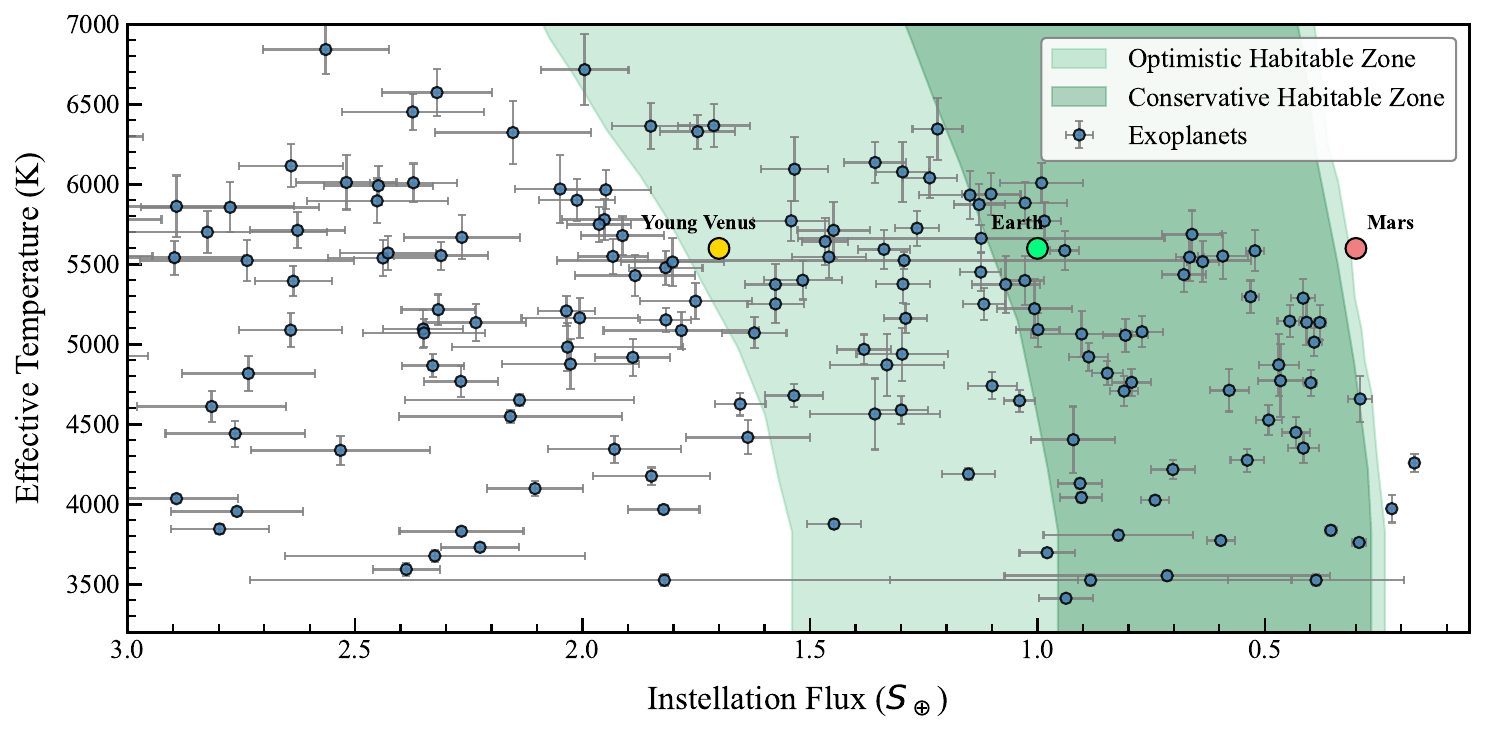}
    \caption{Effective temperature and instellation flux based on the revised planetary parameters. Some extreme outliers mentioned in Section 4 and $S_\oplus$ >3 are excluded. The dark green region shows the optimistic habitable zone, while the light green region represents the conservative habitable zone, based on the assumption that ancient Mars and Venus may have had favorable conditions for sustaining liquid water on their surfaces. However, there remains considerable uncertainty in the instellation flux of some confirmed and candidate planets located near the optimistic habitable zone. Additional observations could help provide a clearer determination of whether these planets fall within or outside that zone. $\sim$20 confirmed and candidate planets reside near the boundaries (inner/outer) of the optimistic habitable zone.} 
    \label{fig:hz}
\end{figure*}

\subsection{Statistical Validation}
\label{sec:triceratops}
While discussing planet catalogs and occurrence rate studies, we need to scrutinize and filter transit signals that are not caused by planets. Reliability is a crucial metric in these contexts, assessing both the individual- and population-level confidence that a signal is of planetary origin. In Kepler data, two distinct classes of phenomena cause signals that are not of planetary origin:

\begin{itemize}
    \item \textbf{Astrophysical False Positives (FP)}:  Signals that resemble on-target planetary transits but are caused by other astrophysical objects, such as eclipsing binary stars, hierarchical triple eclipsing systems, and off-target transiting or eclipsing systems. These false positives can be especially difficult to distinguish from giant planets, which produce similar transit depths as brown dwarfs and grazing stars, and are the dominant phenomena contaminating high-signal-to-noise ratio detections \citep{bryson2020probabilistic,bryson2025estimating}.
    \item \textbf{False alarms (FA)}: These are transit lookalike detections caused by other astrophysical effects (e.g., stellar variability) or non-astrophysical effects (e.g., instrumental noise and systematics). In Kepler data, the major sources of FA are instrumental artifacts, particularly rolling bands and statistical and pixel fluctuations\citep{bryson2020probabilistic}. Rolling bands are thermally dependent electrical signals that move across the CCDs \citep{van2009kepler,caldwell2010instrument}, which produce transit-like signals with a nearly regular ephemeris, especially around the spacecraft's 372-day orbital period. Statistical and pixel fluctuations include unrelated dips caused by cosmic ray hits, or random statistical variations that can accidentally align with a regular ephemeris. These false alarms are common for long-period signals. Planets are especially difficult to distinguish from FAs when the transit signal has a low signal-to-noise ratio. Additionally, false alarm transit detections can arise from stellar variability, which typically occurs for short  periods\citep{bryson2020occurrence}. 
\end{itemize}

False Positive Probability (FPP) and Nearby False Positive Probability (NFPP) quantify the confidence that a detected signal truly originates from a planet rather than from an astrophysical impostor such as an eclipsing binary. Statistical validation refers to the process of statistically ruling out FP scenarios with a sufficient level of certainty to prioritize follow-up observations that can potentially confirm bona fide planets. Typically, statistical validation techniques focus on quantifying FPP and NFPP values for high-S/N candidates, under the assumption that the FA probabilities at high-S/N are negligible \citep{morton2016false,triceratops,bryson2020probabilistic,bryson2020occurrence}.

For statistical validation, we used the publicly available \textit{Bayesian Tool for Rating Interesting Candidate Exoplanets and Reliability Analysis of Transits Originating from Proximate Stars} \citep[TRICERATOPS;][]{triceratops} (version 1.0.21), which computes the FPP and NFPP for each confirmed/candidate transit signal employing the \texttt{TRILEGAL} Galactic model \citep{trilegal...26..165G} to simulate a synthetic stellar population that matches the target star’s characteristics within a cone along its line of sight. At first, \texttt{TRICERATOPS} determines how much flux each nearby star contributes to the photometric aperture and identifies the stars bright enough to produce the observed transit depth. It then compares the provided transit observation with light-curve models of transiting planets and eclipsing binaries to compute the marginal likelihood of different scenarios. Combining these likelihoods with prior probabilities, \texttt{TRICERATOPS} calculates the probability of each scenario, and based on the probabilities, we can classify the planet candidate as a validated planet, likely planet, or likely nearby false positive \citep{triceratops}. The purpose of using \texttt{TRICERATOPS} here is not to determine the status of planets (e.g., candidate or confirmed), but rather to provide an additional, useful, complementary metric to help inform and prioritize follow-up observations and occurrence rate studies.

In \texttt{TRICERATOPS}, we applied a similar detrending procedure, as illustrated in Figure \ref{fig:detrending_workflow}. We applied the default outlier removal ($\sigma=5$) for all planets and masked between 2 to 3 times the transit duration for the phase-folded light curve, depending on the baseline, and applied a small bin size of 0.03 days. As in \texttt{TRICERATOPS}, FPP is expected to have some scatter across runs \citep{triceratops}, so we performed the calculation 20 times for each confirmed planet or candidate and estimated the mean and standard deviation of the resulting distribution. 

\section{Results and discussion}

\label{sec:results}
\begin{figure}[tb!]
\centering
\fig{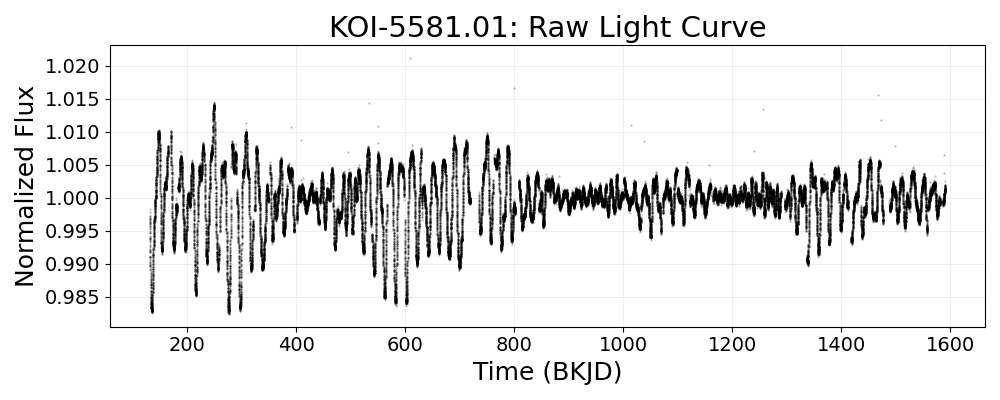}{0.85\columnwidth}{}

\fig{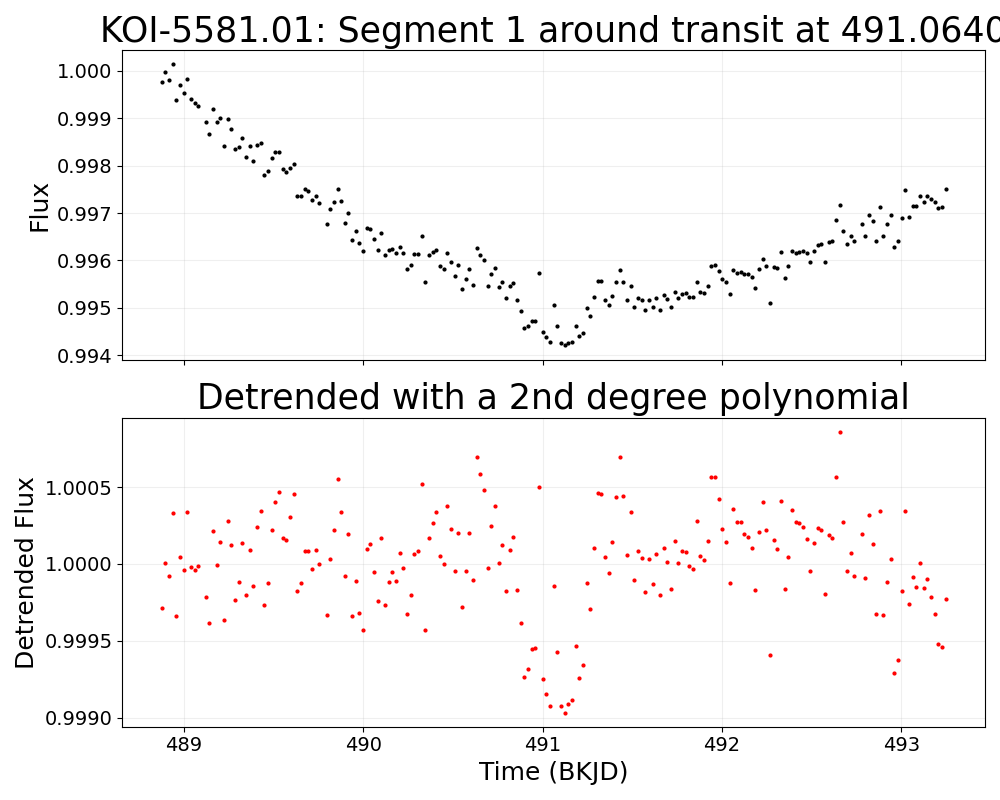}{0.85\columnwidth}{}

\fig{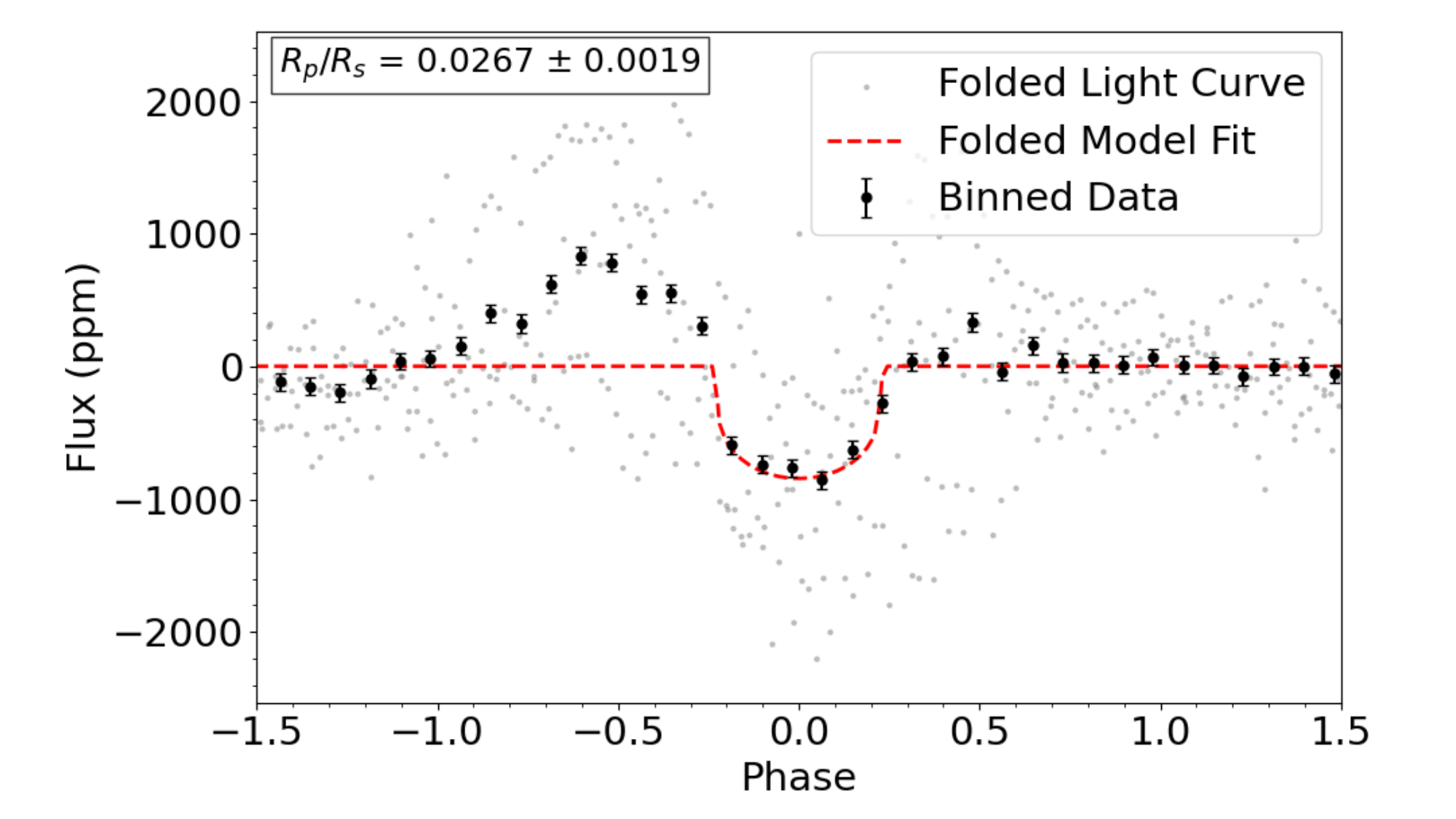}{0.85\columnwidth}{}

\fig{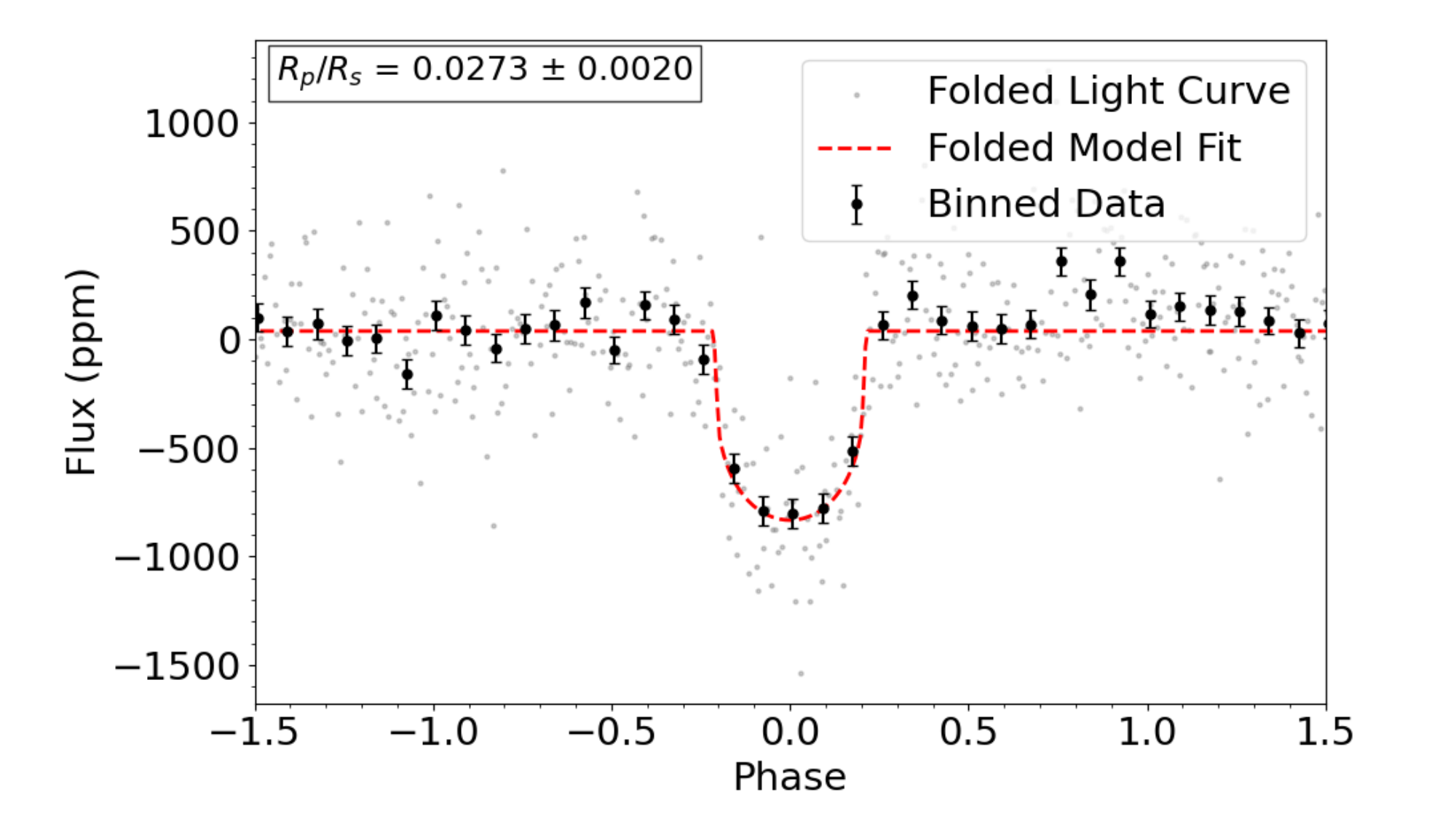}{0.85\columnwidth}{}

\caption{For the planet KOI-5581.01: (a) Shows the Kepler light curve over four years of observations. 
(b) Shows one of the clipped segments of the light curve around the transit, before and after polynomial fitting. 
(c) Shows the fitted phase-folded light curve without clipping or detrending. 
(d) Shows the fitted phase-folded light curve after clipping the light curves and fitting with a second-degree polynomial.}
\label{fig:fitting}
\end{figure}

\subsection{Revised Planetary Parameters}
\label{sec:revised}
The results are presented in Figures \ref{fig:40_transit_fits} through~\ref{fig:fitting}, and the revised parameters are listed in Tables ~\ref{tab:mcmc} and ~\ref{tab:sterna_derived}. Figure 
\ref{fig:40_transit_fits} displays fits for 40 planets and planet candidates selected from the sample of 187. In most cases, our values for planet-to-star radius ratio ($R_p/R_*$), planet distance over star radius ($a/R_*$), orbital period, inclination, and transit epoch ($T_0$) agree with those reported in the Kepler Q1–Q17 DR25 (Table 3).
Owing to our use of updated stellar parameters (Sec.~\ref{sec:stellar}), we find discernible deviations in planet radius ($R_\oplus$) and Instellation flux ($S_\oplus$) relative to Kepler Q1–Q17 DR25 values; some planets are considerably larger or smaller, or closer or farther from the HZ, than previously reported. 

Fig.~\ref{fig:comparison} shows that while our derived values of $R_p/R_*$ and $a/R_*$ are generally consistent with those from DR25, the differences in the uncertainties of those parameters are larger.  In some cases, this is attributable to planets with highly grazing transits (e.g., KOI-1707.02 and KOI-4636.01) whose orbital geometries make them more difficult to reliably characterize and generally less useful for occurrence-rate calculations.
To assess whether our choice of priors (cf.\ Table~\ref{tab:mcmcsettings}) influenced our derived precision on these parameters, 
{We repeated the analysis using substantially broader priors ($0 < R_p/R_* < 0.2$ and $(a/R_*)_{\rm guess} \pm 70\%$) for the 21 objects for which the uncertainties varied by at least a factor of two. We found no significant differences in the resulting $R_p/R_*$ or  $a/R_*$ values; although the latter exhibited a modest shift in the posterior median, it remains statistically consistent at the $<1\sigma$ level.  Thus, the differences in posterior distributions are not heavily sensitive to the width of priors used. A second-order effect may also result from our use of updated stellar parameters, which affect the limb-darkening priors and thus the detailed transit shape, along with ingress and egress. Therefore, differences relative to Kepler DR25 may instead be attributable to some combination of (1) different detrending techniques, (2) different fitting framework and methodology, (3) different limb-darkening values, and (4) our choice of priors.}




{However, the updated stellar radii, masses, and luminosities (see Sec.~\ref{sec:stellar}) directly influence the derived planetary parameters, particularly the planetary radius and the instellation flux. Figure~\ref{fig:comparison_radius} reveals that while most planetary radii are consistent with their DR25 values, some show distinct changes --- these are due to changes in $R_*$ and result in several planets with DR25 radii $< 2 R_\oplus$  now having $R_P \gtrsim 4 R_\oplus$.  For example, KOI-3497.01, KOI-5651.01, and KOI-5652.01 all have stellar (and thus, planetary) radii $>2\times$ larger than originally reported.
{Also, 137 planets exhibited a change in instellation flux $S_\oplus$ of at least $10\%$ relative to their previously reported values, which shifts their location relative to the HZ. 
For example, we now find that KOI-5651.01, KOI-5652.01, KOI-3497.01, KOI-1404.02, KOI-494.01, KOI-2404.01, and KOI-253.02 all receive significantly higher instellation fluxes and should be excluded from any HZ or HZ-adjacent samples. 
However, Figure~\ref{fig:hz} shows that we still see significant uncertainty in the instellation flux of $\sim$20 targets that lie very near the optimistic habitable zone. Further observations could provide a more definitive assessment of their placement within or outside the optimistic habitable zone (see Sec.~\ref{sec:etaearth}). 

\cite{berger26} also derived the radius and instellation flux of thousands of Confirmed and Candidate Planets from the Kepler, K2, and TESS missions. That analysis adopted the NASA Exoplanet Archive's Cumulative KOI table values \citep{cumulative} for estimating those derived planetary parameters. We compared our ${R_p}$ and ${S_\oplus}$ with \cite{berger26} values. The results were mostly consistent. The ${R_p}$ values deviates by $>2 \sigma$ for only 14 systems, while for ${S_\oplus}$ all the values show $<1\sigma$ agreement.}

{For the estimation of $\eta_\oplus$, the most relevant planets are typically considered to lie within the ranges $0.5$--$2.5 R_\oplus$ in radius and $0.2$--$2.2 S_\oplus$ in instellation flux} \citep{bryson2020occurrence,bryson2020probabilistic}. Applying these criteria to our sample, we identify 80 confirmed and candidate planets following our analysis} (Table~\ref{tab:etaearth}). 
We further note that many targets lie very close to the boundaries of our above considered parameter space, with several residing near the selection thresholds; for example, KOI-463.01 ($2.52\pm0.18 R_\oplus$), KOI-2418.01 ($2.22 \pm 0.086 S_\oplus$). A more comprehensive approach to estimating $\eta_\oplus$ would therefore include those excluded borderline Confirmed/candidate planets as well.
Thus, a larger number of targets lie within a few error-bar widths of these parameter limits and so are likely still relevant for $\eta_\oplus$ calculations; nonetheless, the targets in Table~\ref{tab:etaearth} are to be the most relevant for such efforts.


\begin{figure}[tb!]
    \includegraphics[width=1\linewidth]{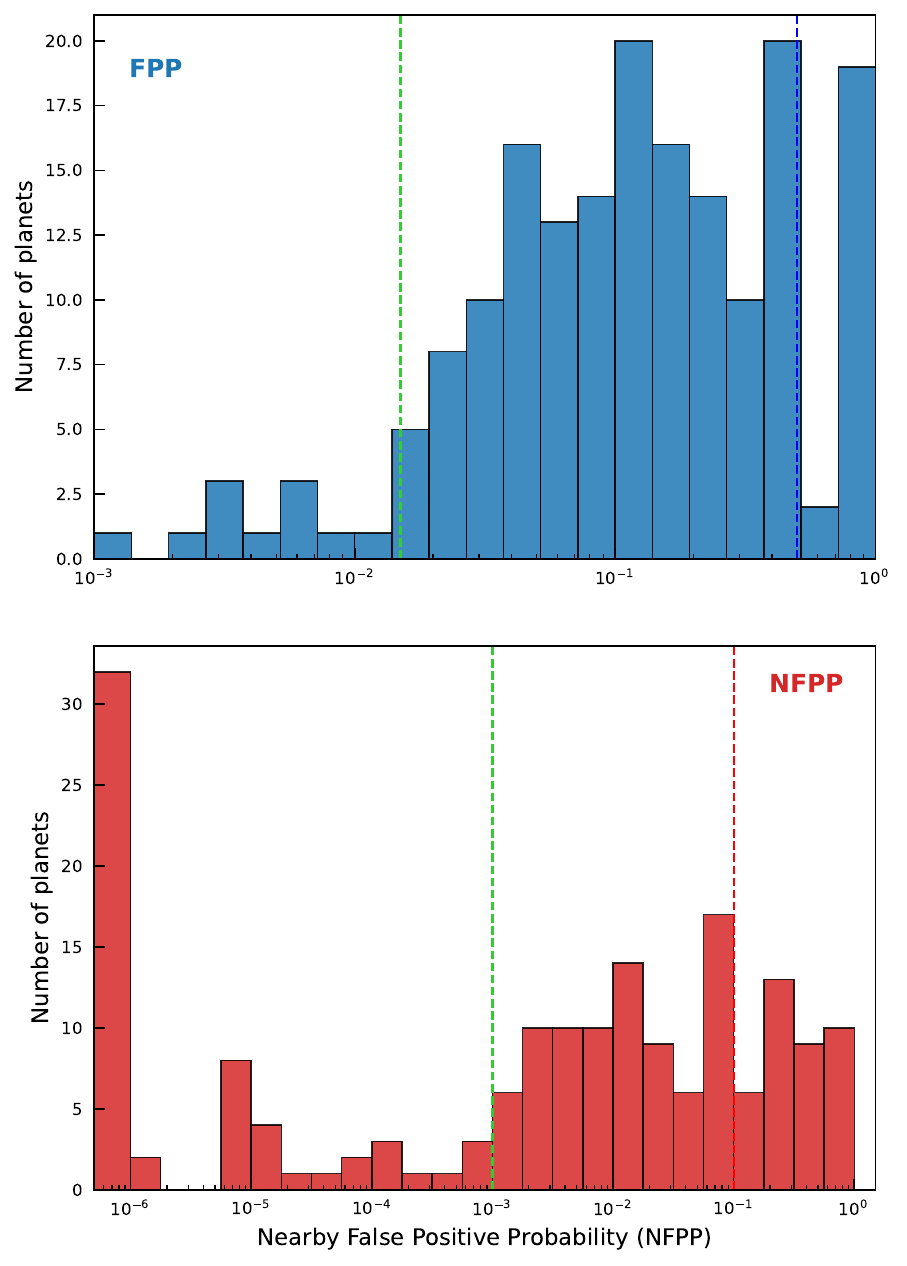}
    \caption{{False positive probability (FPP; top panel) and nearby false positive probability (NFPP; bottom panel) distributions derived using \textsc{TRICERATOPS} (Sec.~\ref{sec:triceratops}). The green dashed vertical line marks the validated-planet regime, defined by $\mathrm{FPP}<0.015$ and $\mathrm{NFPP}<10^{-3}$.  Following the classification scheme of} \cite{triceratops}, the blue dashed vertical line in the top panel denotes the likely-planet regime ($\mathrm{FPP}<0.5$), while the red dashed vertical line indicates the likely false-positive threshold ($\mathrm{NFPP}>0.1$).} 
    \label{fig:FPP_NFPP}
\end{figure}

\begin{figure}[tb!]
    \includegraphics[width=1.05\linewidth]{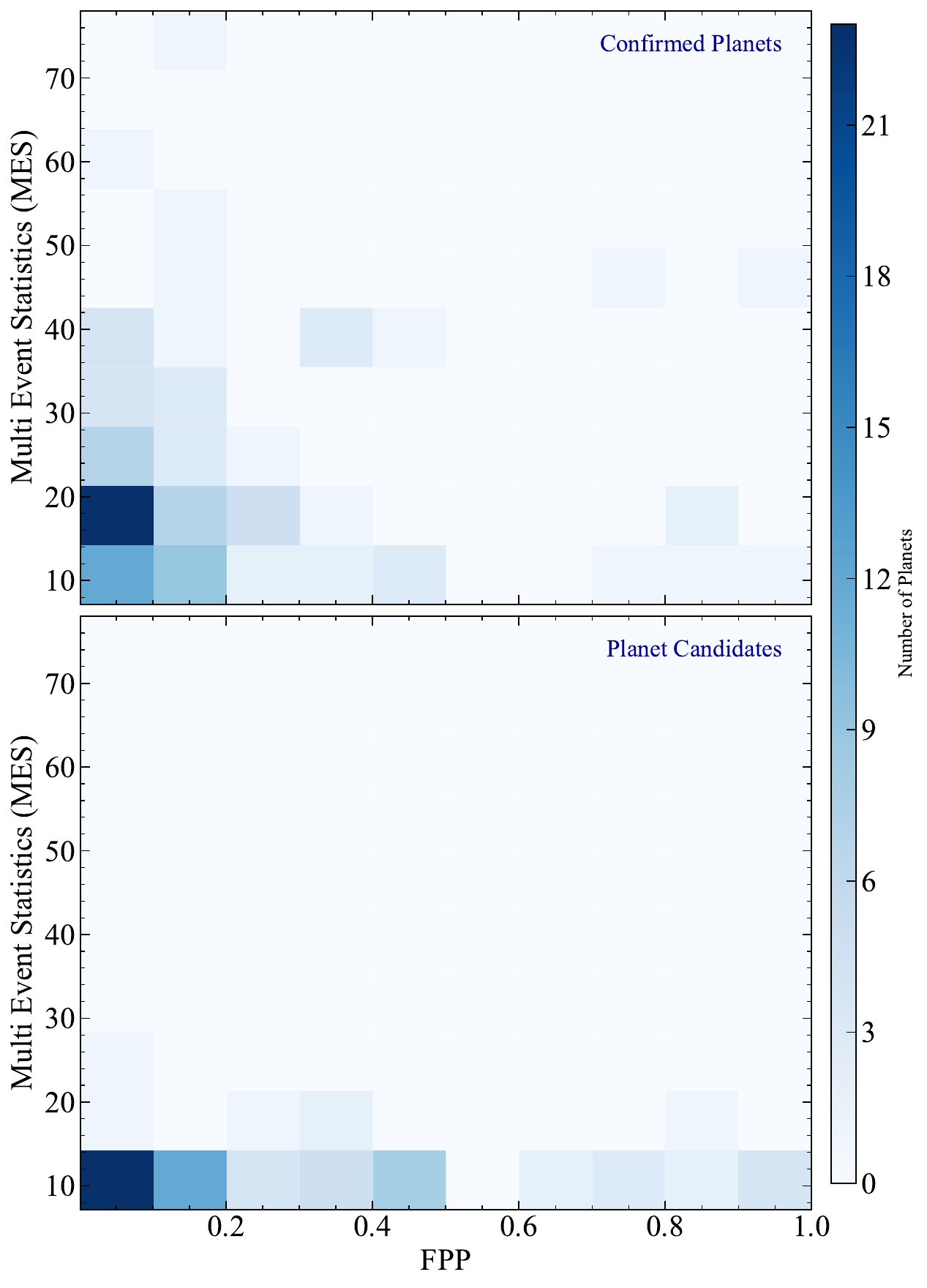}
    \caption{
    Distribution of planets in our sample in terms of False positive probability and Multiple Event Statistics (MES) 
    }
    \label{fig:MES_FPP}
\end{figure}

\subsection{Statistical Validation Results}
\label{sec:stats}
In the statistical validation analysis presented in Table \ref{tab:sterna_derived} and shown in Figures \ref{fig:FPP_NFPP} and \ref{fig:MES_FPP} results could not be obtained for nine planets due to the unavailability of simulated stellar data from the TRILEGAL server \citep{trilegal...26..165G}. \cite{triceratops} classifies confirmed/candidate planets into validated planets 
($\mathrm{FPP} < 0.015$ and $\mathrm{NFPP} < 10^{-3}$), likely planets 
($\mathrm{FPP} < 0.5$ and $\mathrm{NFPP} < 10^{-3}$), and likely nearby false positives 
($\mathrm{NFPP} > 10^{-1}$).
Using that classification scheme, we find that 9 confirmed/candidate planets fall in the statistically validated domain and 43 confirmed/candidate planets in the likely planets domain. 
Furthermore, 37 objects from our list fall into the likely nearby false positive category ($NFPP > 10^{-1}$), 
and the remaining 89 objects lie in intermediate parameter space outside these classification regions. Around $43\%$ (78  planets) yield false positive probabilities (FPP) values $\leq 0.1$ (Figure 8). The overall FPP distribution is skewed toward lower values. The NFPP values remain below $\leq 0.10$ for $\sim78\%$ of the sample (Figure \ref{fig:FPP_NFPP}).


However, 8 of the systems in our sample exhibit relatively higher FPP values ($>0.50$): KOI-581.02, KOI-401.02, KOI-3497.01, KOI-3284.01, KOI-2172.02, KOI-1422.04, KOI-947.01, KOI-2862.01, although these KOIs are listed as Confirmed in the NASA Exoplanet Archive’s disposition. In \cite{triceratops}, the FPP and NFPP values (equations 4 and 5 and Table 1 in \cite{triceratops}) are not the result of only one scenario; rather, they are a combination of 18 different scenarios. In our study for higher values of FPP (for confirmed KOI), the transiting planet probability (TP scenario) from the host star's given parameter constraints was low. 

Additionally, we notice that the Gaia star field used for the false positive analysis contains stars with missing key stellar parameters. In such cases, TRICERATOPS adopts estimated stellar parameters (based on solar values for these sources). In 18 cases in our study, we had to provide the stellar parameters ($R_*$, $M_*$, $T_{eff}$, and parallax) manually (KOIs with asterisks in Table 4). Not only that, but some of the stars (not the Target Star) in the star field were also missing key stellar parameters, which \texttt{TRICERATOPS} handled with solar values. Incomplete or poorly constrained stellar information may affect the characterization of false-positive scenarios, and consequently influence the resulting FPP and NFPP estimates. For these reasons, we recommend evaluating each KOI on a case-by-case basis rather than processing it exclusively through a fully automated pipeline. Whenever missing or invalid stellar parameters are identified, the corresponding values should be manually verified and, where possible, supplemented using reliable catalog or literature sources before performing the false positive analysis.


In Kepler DR25, the FPP values were estimated using \texttt{Vespa} \citep{morton2016false}. For our sample, DR25 has comparatively lower FPP values (closer to Zero) than our estimates. Our FPP estimates with \texttt{triceratops} show a wide range of differences between the two tools. For 27 confirmed/candidate planets (16.1\%), the \texttt{TRICERATOPS} FPP estimates were lower than the DR25 FPP values. 21 of the confirmed/planet candidates (12.5\% of the total sample) differed by less than a factor of three. The remaining 120 confirmed/planet candidates (71.4\%) differed by more than a factor of three. However, we note that the use of \texttt{Vespa} is no longer recommended for statistical validation studies \citep{remove_vespa}.

The  Multiple Event Statistic (MES) is effectively the signal-to-noise ratio (SNR) obtained by combining multiple periodic transit signals detected in a {\em Kepler} light curve. For Kepler, a candidate planet detection requires MES $\geq 7.1$, a threshold considered by the \texttt{Robovetter} \citep{coughlin2020robovetter} in the DR25 catalog \citep{DR25}. While no strong separation between confirmed and candidate planets is present in the FPP distribution, planets with higher MES generally exhibited lower FPP values (Figure~\ref{fig:MES_FPP}), in agreement with trends identified by \cite{triceratops,morton2016false}.

However, the number of confirmed and candidate planets identified within the statistically validated regime is relatively small and is therefore unlikely, on its own, to significantly advance occurrence-rate studies. A more comprehensive approach would be to also include planets in the likely-planet regime and incorporate additional metrics, such as reliability, the availability of follow-up observations, and other relevant factors when performing such analyses. The ambiguity associated with systems having conflicting FPP and NFPP values (i.e., statistical validation metrics) can also be resolved through follow-up observations.

\begin{figure}[ht!]
\centering

\fig{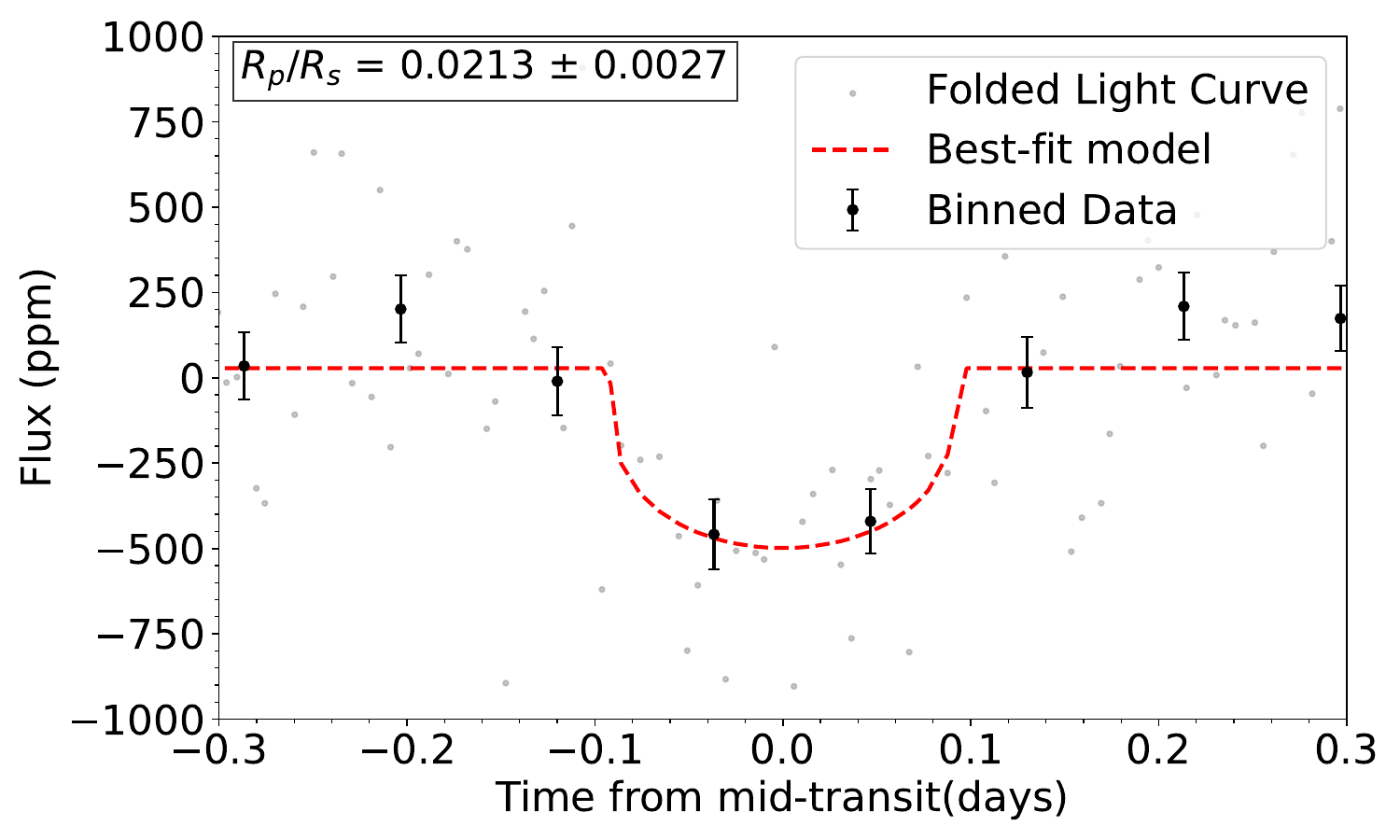}{0.9\columnwidth}{(a)}

\fig{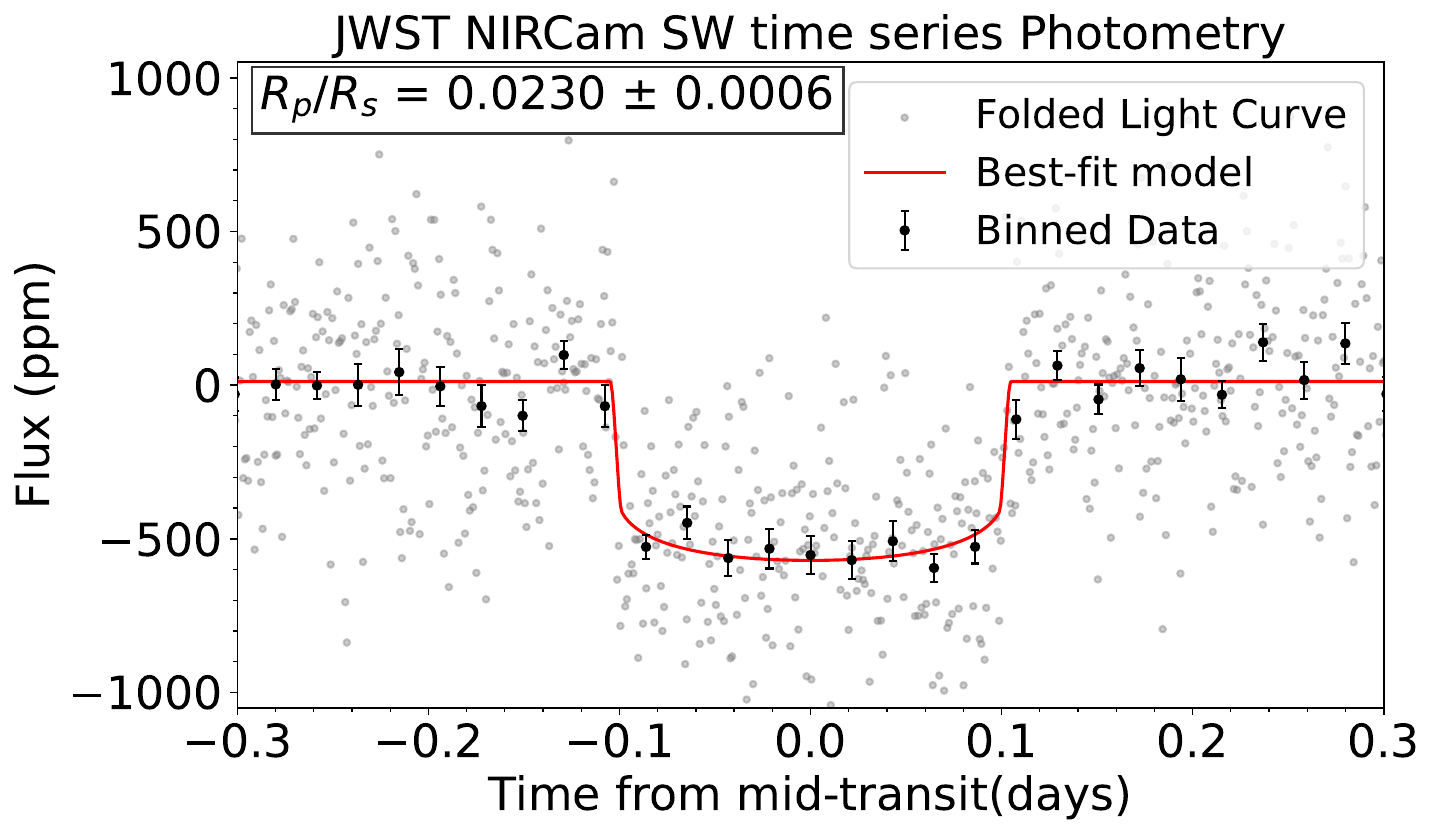}{0.9\columnwidth}{(b)}

\fig{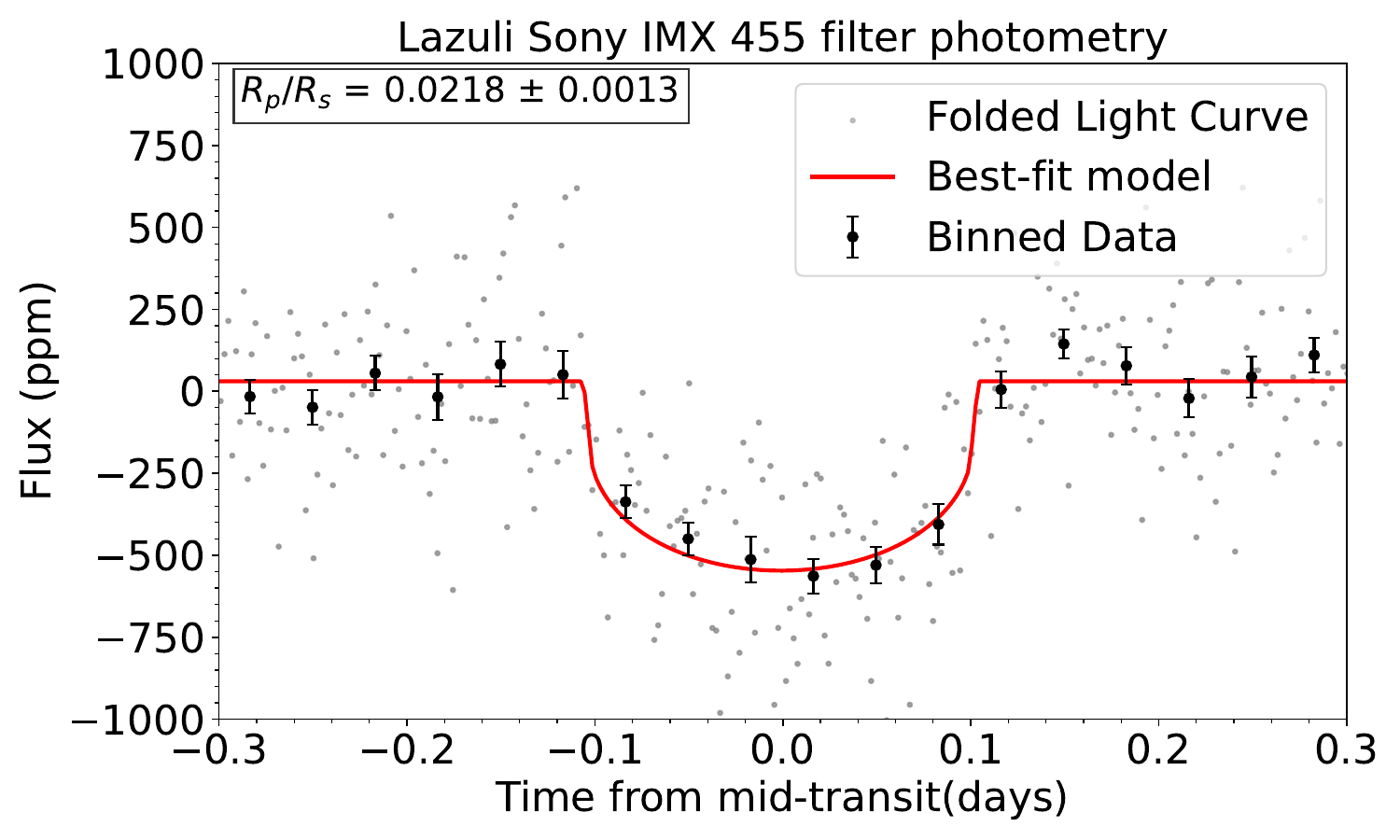}{0.9\columnwidth}{(c)}

\caption{For planet candidate KOI-8047.01 (a) Shows Kepler analysis for four years of observation 
(b) Shows simulated JWST NIRCAM SW time series photometry analysis B. 
(c) Shows simulated Lazuli 3m telescope WCC broadband photometry}
\label{fig:vertical_three_sim}
\end{figure}

\begin{figure}[ht!]
    \centering
    \includegraphics[width=1.0\linewidth]{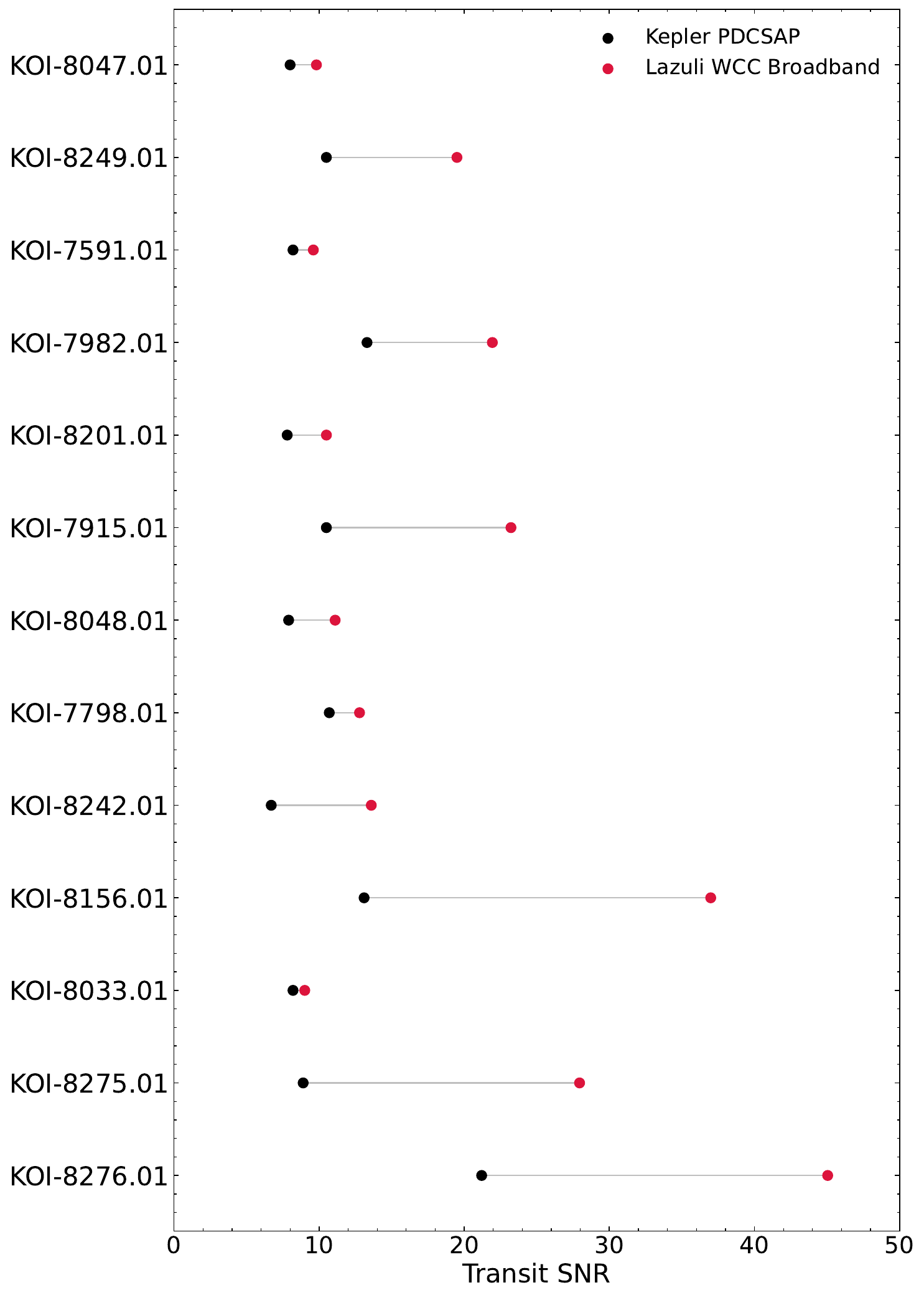}
    \caption{{Transit SNR comparison between one simulated transit with Lazuli WCC Broadband photometry and our measurements of four full years of Kepler photometry (Table 3) for candidate planets with 300\,d$<P<$400\,d. Follow-up of just a single large-aperture transit improves the transit SNR by roughly 1.8$\times$.}}
    \label{fig:comparison_transit}
\end{figure}

\subsection{Discussion and Next Steps}

Constraints on the planetary parameters, statistical validation, and reliability estimates of these confirmed/ planet candidates can be significantly improved by revisiting their transits one or more times. From Table~\ref{tab:etaearth}, we can see that several of these objects lack sufficient follow-up observations, including RV and transit observations, to be placed in the confirmed category and have ambiguous FPP/NFPP values. To enhance the reliability of these detections and enable more robust statistical validation, additional observations with the James Webb Space Telescope (JWST) \citep{beichman2014observations,rigby2023science}, PLATO \citep{rauer2016plato}, Lazuli Space Observatory \citep{roy2026lazuli}, Earth 2.0 (ET)\citep{2024ChJSS..44..400G}, or other space-based observatories could provide improved constraints on their planetary nature and parameters.

Kepler had a 0.95 m aperture; observations with larger-aperture telescopes can provide better per-transit photometric precision. In addition, the finer angular resolution of these facilities enables better isolation of the host star on the sky, reducing the effective region that must be searched for background sources in statistical validation analyses. The improved photometric precision and finer angular resolution together yield a higher signal-to-noise ratio (S/N) of individual transits, thereby refining or further constraining the planetary parameters and statistical validation metrics. {This is particularly crucial for planets near the Kepler detection threshold ($7<\mathrm{MES}<25$) and candidates with an Earth-like orbital period ($P>300$). Follow-up observations can also reduce and even entirely remove the ambiguity in the statistical validation metrics discussed in Section~\ref{sec:stats}.
}

To investigate this, we simulated the transit light curve of Candidate Planet KOI-8047.01 (reliability = 0.73 \citep{bryson2020occurrence}, FPP = 0.87, and NFPP=0.38) as observed by JWST and the upcoming 3.1 m Lazuli Space Observatory  \citep{roy2026lazuli}. {KOI-8047.01 is listed in Table~2 and is relevant to $\eta_{\oplus}$ calculations} \citep{bryson2020occurrence}.  

For JWST, we simulated observations with the NIRCam short-wavelength time-series photometry mode with CLEAR SW Pupil and F150W2 filter \citep{beichman2014observations}. For Lazuli, we used an early internal version of the Widefield Context Camera (WCC) using a Sony IMX455 sensor \citep{roy2026lazuli} with a broadband filter (300-1100 nm) to generate proof-of-concept simulations. This version does not yet include a full end-to-end treatment of the observatory. We simulated 15 hours of time-series data for JWST (1.22 min cadence), while for Lazuli, we simulated time-series observations (2.8 min cadence) extending 10 hours before and after the transit. 

Figure~\ref{fig:vertical_three_sim} presents a comparison between the Kepler photometry, the JWST NIRCam short-wavelength time-series photometry, and the Lazuli photometry. For KOI-8047.01, we find that JWST and Lazuli achieve comparatively better precision than Kepler’s four years of observations, although we note that the Lazuli WCC ETC and STERNA do not include a full end-to-end treatment of the observatory's properties. 


{Since Lazuli, with a 3.1-meter aperture, is expected to precisely characterize confirmed and candidate exoplanets with long orbital periods ($>300$ days), it is expected to achieve significantly higher transit signal-to-noise ratios (SNRs).


To illustrate this, we simulate 12 additional candidate planets orbiting F, G, and K-type stars using \textsc{STERNA}. The majority of them are from Table ~\ref{tab:etaearth}, and all these systems have Earth-like orbital periods exceeding 300 days and less than 400 days. In all cases, we obtain higher transit SNRs compared to the Kepler DR25 values, as shown in Figure ~\ref{fig:comparison_transit} We find a maximum SNR improvement of $\sim 3\times$, and the mean improvement is $\sim 1.8\times$ with a single transit.

\subsection{Toward an Improved Value of Eta-Earth}
\label{sec:etaearth}
{Our new catalog of revised planet and system parameters provides a useful foundation for future efforts to refine estimates of $\eta_\oplus$, a quantity particularly sensitive to such updates} \citep{fernandes2025we,bergsten2026modeling}. Although small sample size may be the limiting factor in such studies \citep{bryson2020occurrence}, improved planet parameters and reliability can help improve $\eta_\oplus$ ``around the edges'' until a larger HZ and near-HZ sample can be compiled \citep{bryson2025estimating}.  

\subsubsection{The Benefit of Large Aperture}
The reason that space-based transits from JWST or Lazuli could dramatically improve transit SNR is their significantly greater collecting area compared to Kepler.  Kepler operated for four years, so a planet with orbital period $P$ was observed to transit roughly (4\, yr)/$P$ by {\em Kepler}. For a planet with Kepler-derived transit  $\mathrm{SNR}_\mathrm{Kep} \approx \mathrm{MES}$, in the photon-noise limit, another observatory with identical throughput and bandpass but primary mirror diameter $D$ will observe a per-transit SNR of
\begin{equation}
    \mathrm{SNR}_\mathrm{Tran} \approx \left( \frac{D}{0.95\,\mathrm{m}} \right)  \left( \frac{P}{4\,\mathrm{yr}} \right)^{1/2}  \ \mathrm{SNR}_\mathrm{Kep} 
\end{equation}
Thus for the longest-period Kepler systems, a single photon-limited, space-based transit with a large-aperture observatory could roughly double the transit SNR and consequently improve the precision on derived parameters; this factor-of-two improvement is consistent with the average improvement factor of 1.8 discussed in the preceding section and shown in Fig.~\ref{fig:comparison_transit}.  Doubling the precision on transit depth implies a commensurate two-fold improvement on the precision of $R_P/R_*$, which would allow a significant improvement over current measurement precision. With a 3\,m aperture, a single transit  should be at least detectable (i.e., SNR$_\mathrm{Tran}\gtrsim\mathrm{SNR}_\mathrm{Kep}$) for $P \gtrsim 140$\,d.   Furthermore, an independent detection of the transit from a fundamentally different observatory, even if only at moderate S/N, would help address False Alarm estimates as well as} increase the likelihood of the transit being real and not merely a data artifact.  In addition,  if the observations were acquired with higher angular resolution than was permitted by {\em Kepler}'s 4'' pixels (as would be feasible with both JWST and Lazuli) then the ability to identify false positives resulting from blended eclipses of faint stars near the target system would also be significantly enhanced.  Thus the kind of transit follow-up showcased in Fig.~\ref{fig:comparison_transit} could improve derived planet parameters, increase statistical reliability of the signal, and ultimately contribute to improved estimates of  $\eta_\oplus$ and related demographic quantities.

A next step in improving the uncertainties on $\eta_\oplus$ using the {\em Kepler} sample is therefore to derive a quantitative estimate of how the uncertainties in $\eta_\oplus$ can be improved using techniques and facilities now (or soon) available. This includes (a) the effect of well-determined candidate reliabilities on $\eta_\oplus$ , since new observations should result in less ambiguity as to whether or not transit signals are merely false positives; and (b) the effect of well-measured system parameters, since improved planet radii determine the likelihood a planet is rocky and improved insolation values determine the likelihood that the planet is in the HZ.  In addition to these possibilities (which rely largely on additional, precise transit photometry of HZ and near-HZ candidates) there is also (c) the effect of additional supporting ground-based observations (e.g., spectroscopically-derived stellar parameters or high-resolution imaging) on false-positive probabilities and associated candidate reliabilities.

Although points (a) and (b) above may provide the most scope for near-term improvements to $\eta_\oplus$, they will likely require relatively expensive space-based observations (see discussion below).  Point (c) is more easily addressed.  By consulting the ExoFOP\footnote{\url{https://exofop.ipac.caltech.edu/tess}} we find that $\sim$60\% of our sample still lacks high-quality (resolution $<$0.1'') high-resolution imaging necessary to rule out most blends; luckily, this lack will soon be remedied by the Roman-Kepler Legacy Survey\footnote{{\em Roman} Space Telescope General Astrophysics Survey Program 2004, PI Kraus; \url{https://vmromanweb1.ipac.caltech.edu/cycle1-approved-programs/2004}}.  Many spectroscopic follow-up efforts have focused on {\em Kepler} candidates,  perhaps the deepest and most uniform of which was the Keck/HIRES California-Kepler Survey \citep{petigura:2017}; however, $\sim$40\% of our sample still lacks   HIRES spectra (and many still lack any recorded stellar spectroscopy). Finally,  while our sample's stellar parameters are unlikely to change further given our incorporation of Gaia DR3 stellar parameters, the use of multi-year  Gaia (DR4)  astrometric data may also be useful in identifying previously unknown stellar binaries among our sample.

\subsubsection{The Benefit of Improved Reliability}

Reliability for Kepler Confirmed/Candidate planets has two distinct components: reliability against astrophysical false positives, estimated from $R_{\rm FP}=1-\mathrm{FPP}$, and reliability against false alarms, $R_{\rm FA}$. As discussed in \ref{sec:triceratops}, $R_{\rm FP}$ quantifies the probability that a signal is not produced by an astrophysical false-positive scenario, whereas $R_{\rm FA}$ quantifies the probability that it is not produced by instrumental systematics, stellar variability, or other false-alarm sources. The total reliability is therefore,
\[
R = R_{\rm FP}R_{\rm FA} = (1-\mathrm{FPP})R_{\rm FA}
\]
\citep{bryson2020probabilistic,bryson2020occurrence}. In the following analysis, $R$ denotes this combined reliability.  \citet{bryson2020occurrence} showed that occurrence-rate estimates from different Robovetter threshold catalogs are inconsistent without reliability correction, but become mutually consistent once completeness and reliability are both included. 


We adopted the available DR25 based $R_{FA}$ values from \cite{bryson2020occurrence,bryson2020probabilistic}, and the publicly available GitHub repository \footnote{\url{https://github.com/stevepur/DR25-occurrence-public}} and estimated the total reliability using the FPP values from TRICERATOPS. The reliability values are provided in Table 4. In our study, we found that 31 lie at the high-reliability edge ($R \geq 0.95$), 2 at the low-reliability edge ($R < 0.05$), and the remaining 107 in the intermediate region. In the intermediate space, the mean reliability is 0.68, and the median reliability is 0.73, and 18 of them have reliability $>0.90$.

These intermediate systems are the most natural targets for follow-up, since new high-precision observations from Lazuli, JWST, or similar facilities could help clarify whether the signals are planetary, astrophysical false positives, or false alarms.

We characterize the catalog-level reliability ambiguity as $A = \sum R_i(1 - R_i)$, which is maximized for candidates near R = 0.5 and vanishes as the reliability approaches either extreme; the intermediate population carries 92\% of the total ambiguity in our sample. Because a sufficiently precise follow-up observation drives a candidate's reliability toward one extreme or the other, the ambiguity removed by observing a chosen set of targets is simply $\sum R_i(1 - R_i)$ over that set, independent of the direction in which each candidate’s reliability shifts: the information gain from follow-up is guaranteed in advance. A single observation of a candidate near R = 0.5 removes $\sim$0.25 from A, the maximum achievable per target. Prioritizing the most ambiguous systems, follow-up of 5, 10, and 20 candidates reduces A by approximately 14\%, 27\%, and 53\%, respectively.
In contrast, the direction of the change in the reliability-weighted effective planet count, $\sum R_i$, depends on the outcomes: the ten most ambiguous candidates in our sample (0.36 $\leq$ R $\leq$ 0.56) would raise the effective count by 4.9 planets if all are ultimately classified as planets, or lower it by 5.1 if all are ultimately classified as false positives. If the current reliabilities are well-calibrated probabilities, each candidate is planetary with probability $R_i$, so the expected net change in $\sum R_i$ for this set is approximately zero. The principal benefit of high-precision follow-up is therefore not an increase in the mean reliability of the catalog but the elimination of its ambiguity, $\Delta A$. While reliability uncertainty has limited effect on marginalized occurrence rates \citep{bryson2020probabilistic,bryson2020occurrence}, follow-up observations test the reliability calibration itself. We emphasize that the ambiguity $\Delta A$ removed by follow-up is agnostic to technique: any observation that decisively shifts a candidate's reliability contributes, so precise space-based transit photometry, high-resolution imaging that rules out (or reveals) blended stellar companions, and spectroscopy that identifies binarity or refines stellar parameters (point (c) above) each remove the same per-target ambiguity. We further note that the same single-transit observations (Fig. \ref{fig:comparison_transit}) simultaneously improve the transit-depth precision of these targets, so that one observation can better constrain both a candidate’s reliability and, as we show below, its membership in the $\eta_\oplus$ parameter domain.

The improvement in transit SNR is directly relevant to the boundary cases that enter occurrence-rate calculations, because the inclusion of a planet in the adopted $\eta_\oplus$ domain depends not only on its nominal parameters but also on their uncertainties.
Inclusion may be treated either probabilistically, according to the fraction of a planet's posterior that lies within this domain, or as a binary classification.
Using the nominal values alone, we identify 80 planets in our sample within $0.5 < R_p/R_\oplus < 2.5$ and $0.2 < S/S_\oplus < 2.2$.
To quantify the effect of improved precision beyond this nominal-value selection, we adopt a conservative binary criterion requiring the full $1\sigma$ radius--instellation interval to remain inside the domain.
To isolate the effect of improved transit photometry, we hold the nominal planetary parameters fixed and rescale only the radius uncertainties, $\sigma_{R,\mathrm{used}} = n\sigma_R$, while leaving the instellation uncertainties unchanged. Because the experiment rescales the total radius uncertainty, $n$ represents an effective precision scenario; in practice, uncertainty in $R_*$ sets a floor on improvements from transit photometry.
We leave the instellation uncertainties unchanged because they are dominated by the stellar luminosity and mass rather than by the transit photometric precision.
The \(n=0.5\) case is representative of a Lazuli-like single-transit follow-up: the mean simulated SNR improvement of \(\sim 1.8\times\) corresponds to an effective radius-uncertainty reduction of approximately one half when combined with the existing Kepler constraint.
In this case, the number of planets satisfying the \(1\sigma\) \(\eta_\oplus\)-domain inclusion criterion increases from 62 to 70, while the boundary overlapping sample decreases from 38 to 25.
A JWST-like high-precision case, represented by \(n=0.25\), increases the number of included planets to 74 and reduces the boundary overlapping sample to 17.
Thus, even without shifting the nominal planetary parameters, improved radius precision alone can resolve the inclusion status of a substantial fraction of the systems near the adopted $\eta_{\oplus}$-domain boundaries. The two scenarios add 8 and 12 high-confidence members to the current 62-candidate sample, corresponding to increases of approximately 13\% and 19\%, respectively. These changes are substantial relative to the small sample entering an occurrence-rate analysis, although they should not be interpreted as equal fractional changes in $\eta_{\oplus}$, which also depends on detection completeness, reliability, and the individual statistical weight of each system.

\section{Conclusion}
\label{sec:conclusion}
In this paper, we reanalyzed {\em Kepler} transit photometry of 187  planets located in or near the HZ.  The fundamental differences between this and previous analyses are (1) a larger total target list than previous such analyses (see Sec.~\ref{sec:my_sample}), since our targets are drawn from multiple sources;
(2) a uniform, homogeneous analysis of {\em all} these systems, not all of which were included in {\em Kepler}'s DR25; (3) the incorporation of new, updated stellar parameters (see Sec.~\ref{sec:stellar}), which results in slight changes to stellar limb-darkening and more significant changes in the planets' derived sizes and incident insolation levels;  (4) the derivation of statistical false-positive probabilities (see Sec.~\ref{sec:triceratops}), providing a new, independent metric for comparison with the {\em Kepler} project's reported candidate Reliability (and other similar relevant metrics); and (5) an initial, quick look at how sample Reliability and Ambiguity can be improved via follow-up transit observations (see Sec.~\ref{sec:etaearth}).

The results are shown in Figures \ref{fig:40_transit_fits}–\ref{fig:hz} and \ref{fig:FPP_NFPP}-\ref{fig:MES_FPP}, with revised planetary parameters provided in Tables \ref{tab:mcmc} and \ref{tab:sterna_derived}. Overall, the parameters derived with our STERNA tool -- planet-to-star radius ratio ($r_p/R_*$), scaled semi-major axis ($a/R_*$), orbital period, inclination, and transit epoch ($T_0$) -- are largely consistent with those reported in the Kepler Q1–Q17 DR25 catalog (Table 2).
However, several planets exhibit notable shifts in their properties, including $R_\oplus$ and $S_\oplus$, leading to reclassification in terms of size and proximity to the HZ. These revisions stem primarily from updated stellar radii and luminosities, which propagate directly into the relevant derived planetary characteristics. Collectively, the revised parameters and validation insights presented here help refine our understanding of the demographic landscape of potentially habitable exoplanets. Fig.~\ref{fig:hz} shows our entire sample relative to the HZ: most of our targets are within the Optimistic HZ, though Fig.~\ref{fig:comparison_radius} shows that not all these systems have radii consistent with terrestrial planets.  

Our statistical validation analysis (Sec.~\ref{sec:triceratops}) finds that only nine objects satisfy the adopted TRICERATOPS statistical validation thresholds for astrophysical false positive scenarios. However, only five have false positive probabilities of 100\%; Fig.~\ref{fig:FPP_NFPP} shows that most fall in an intermediate regime, with median FPP$\sim$0.1 and NFPP$\sim$0.01.  Although higher MES values appear to have some influence on lowering FPP, no strong correlation is evident (see Fig.~\ref{fig:MES_FPP}). About $15\%$ of the systems exhibit high FPPs despite having relatively large MES values, indicating that they warrant closer examination -- particularly those still classified as Candidates in the Exoplanet Archive Disposition -- due to the potential for contamination or unresolved systematics. Nevertheless,  {\em Kepler}'s noise properties mean that at lower  SNR (e.g., MES $\lesssim 10$), validating transit signals with absolute confidence is often infeasible \citep{mullally2018kepler}. This challenge is especially acute for the longest-period ($\gtrsim$200 days) candidates, where just a few transits were observed by {\em Kepler}. 

We also note that our reported FPP and NFPP values are not intended to be the final word on these HZ and near-HZ planet candidates: rather, we aim to provide another, independent metric for considering these systems in future analyses. There is a strong need for further cross-validation across various validation methodologies and frameworks beyond {\em Kepler}'s RoboVetter and \texttt{TRICERATOPS} (used here) --  including traditional machine-learning models \citep{armstrong2021exoplanet}, and deep neural network (DNN)–based approaches like ExoMiner \citep{valizadegan2022exominer}. Such comparative studies could help to further strengthen our confidence in the planetary status of HZ and near-HZ candidates. Also, using MOLUSC \citep{molusc} with \texttt{TRICERATOPS} can help rule out stellar false positives. This is particularly crucial for constraining the range of properties of unseen companions using astrometric, imaging, and velocity information.


In Sec.~\ref{sec:etaearth} we discuss several possible next steps for improving our analysis here, and for taking the next steps toward deriving an improved estimate of $\eta_\earth$ and related planet-occurrence rates.  Aside from completing the sample of high-resolution imaging and spectroscopy of our targets, the best possibility in the near term is conducting follow-up transit photometry of these candidates with large-aperture space telescopes in order to precisely locate candidates relative to the HZ and the Radius Valley, and to reduce the ambiguity as to whether or not these signals truly arise from planetary transits. Two strong contenders are JWST and the Lazuli space telescopes, which could significantly improve the precision on transit and planetary parameters. Our updated transit fits provide revised ephemerides and associated uncertainties that can be used to assess the feasibility of future transit observations.


Looking ahead, the inventory of habitable-zone planets is expected to expand further with upcoming missions like PLATO, Earth 2.0 \citep{2024ChJSS..44..400G}, and perhaps Roman's  Galactic Bulge Time Domain Survey \citep{gaudi2022roman,2023ApJS..269....5W}. 
With a sufficiently large sample of HZ systems, it will become possible to derive a more accurate $\eta_\oplus$ \citep{bryson2025estimating,fernandes2025we} --- as well as to investigate important factors like stellar and planetary multiplicity\citep{offner..534..275O,Raghvan,huber2016impact,lissauer2024updated,he2019architectures,he2020architectures}, the chemical composition of host stars \citep{mulders2016super,2018AJ....155...89P,lee2019boundary,zhu2024metallicity}, the influence of nearby large planets \citep{2025AJ....170..210G}, and the effects of galactic population, clustering, and environment\citep{winter2020stellar,mustill2022hot,longmore2021impact}. Also, the \textit{Roman-Kepler Legacy Survey} can potentially validate some of the small rocky planet candidates that \textit{Kepler} observed.  

\section{Acknowledgements}
This work is funded by the Astrophysics and Space Institute, Schmidt Sciences LLC. We thank the reviewer for the detailed, and helpful comments that significantly improved the paper. We thank Nicole Arulanantham, Guðmundur K. Stefánsson, Everett Schlawin, Steve Bryson and Steven Giacalone for their invaluable comments and advice on the manuscript. We also acknowledge Ewan Douglas, Pierre Nicholas, Patrick Ingram, Justin Hom, Sanchit Sabhlok, Doug Kelly, Arron Goldtooth and Jonathan Brande for coding support and helpful discussions.
\newpage

\setlength{\tabcolsep}{3pt}
\startlongtable
\begin{deluxetable*}{cccccccccc}
\tabletypesize{\scriptsize}
\tablecaption{Confirmed/Candidate planets relevant for $\eta_\oplus$ estimates after our analysis}
\label{tab:etaearth}
\tablewidth{0pt}
\tablehead{
  \colhead{Planet ID}&
  \colhead{Period (days)}&
  \colhead{$R_\oplus$} &
  \colhead{$S_\oplus$} &
  \colhead{Disposition}&
  \colhead{Reliability (DR25)}&
  \colhead{FPP*}&
  \colhead{NFPP*}&
  \colhead{Reliability*}&
  \colhead{\begin{tabular}{@{}l}$\sigma_{T_0}$\\(Hr) in 2027\end{tabular}}
}
\startdata
KOI-3255.01&$66.65\pm0.00044$&$1.59\pm0.11$&$2.16\pm0.24$&Confirmed&--&$0.141$&$5.00e-02$&$0.859$&$1.02$\\
KOI-7223.01&$317.09\pm0.0067$&$2.19\pm0.19$&$2.15\pm0.17$&Confirmed&--&$0.375$&$3.31e-01$&$0.476$&$3.24$\\
KOI-2529.02&$64.00\pm0.00052$&$2.29\pm0.21$&$2.14\pm0.25$&Confirmed&0.96&$0.144$&$4.61e-03$&$0.850$&$1.28$\\
KOI-812.03&$46.18\pm0.00023$&$2.26\pm0.12$&$2.10\pm0.10$&Confirmed&1&$0.091$&$3.37e-02$&$0.909$&$0.77$\\
KOI-7932.01&$502.25\pm0.025$&$2.12\pm0.20$&$2.05\pm0.098$&Confirmed&--&$0.395$&$3.02e-01$&$0.471$&$7.87$\\
KOI-2290.01&$91.50\pm0.00053$&$1.86\pm0.12$&$2.03\pm0.15$&Candidate&1&$0.148$&$0.00e+00$&$0.852$&$0.91$\\
KOI-2162.02&$199.67\pm0.0038$&$1.33\pm0.090$&$2.01\pm0.083$&Confirmed&0.99&$0.154$&$1.63e-03$&$0.789$&$3.09$\\
KOI-1989.01&$201.11\pm0.0064$&$1.89\pm0.14$&$1.96\pm0.070$&Candidate&1&$0.158$&$0.00e+00$&$0.841$&$4.96$\\
\enddata
\tablecomments{
*FPP, NFPP, Reliability from this work, 
**Could not be processed with Triceratops,
$--$ DR25 total reliability , False alarm reliability $R_{FA}$ not available. \citep{bryson2020occurrence}
}
\end{deluxetable*}
\clearpage

\newpage
\startlongtable
\begin{deluxetable*}{lcccccc}
\tabletypesize{\scriptsize}
\tablecaption{STERNA transit fit results}
\label{tab:mcmc}
\tablewidth{\textwidth}
\tablehead{
  \colhead{Planet ID} &
  \colhead{$r_p/r_*$} &
  \colhead{$a/r_*$} &
  \colhead{Inclination (deg)} &
  \colhead{Transit Epoch ($T_0$, BKJD)}&
  \colhead{Period (days)}&
  \colhead{$\sigma_{T0}$(hr) in 2027 }
}
\startdata
KOI-4636.01&$0.059\pm0.016$&$41\pm5$&$88.60\pm0.17$&$170.98\pm0.021$&$122.75\pm0.0028$&$3.57$\\
KOI-463.01&$0.044\pm0.0011$&$35\pm2$&$88.58\pm0.11$&$185.27\pm0.0010$&$18.48\pm0.000023$&$0.20$\\
KOI-87.01&$0.021\pm0.0018$&$190\pm39$&$89.76\pm0.088$&$133.70\pm0.0035$&$289.86\pm0.0019$&$1.08$\\
KOI-401.02&$0.030\pm0.0018$&$159\pm29$&$89.71\pm0.090$&$251.28\pm0.0064$&$160.02\pm0.0013$&$1.25$\\
KOI-505.05&$0.033\pm0.0025$&$53\pm17$&$89.05\pm0.37$&$187.33\pm0.0053$&$87.09\pm0.00055$&$1.05$\\
KOI-1871.01&$0.038\pm0.0019$&$58\pm9$&$89.15\pm0.17$&$177.44\pm0.0037$&$92.73\pm0.00041$&$0.69$\\
KOI-5790.01&$0.026\pm0.0020$&$150\pm36$&$89.68\pm0.12$&$229.76\pm0.015$&$178.26\pm0.0048$&$4.15$\\
KOI-2210.02&$0.038\pm0.0031$&$167\pm40$&$89.81\pm0.17$&$300.72\pm0.0075$&$210.63\pm0.0022$&$1.63$\\
\enddata
%
\end{deluxetable*}
\clearpage
\newpage
\setlength{\tabcolsep}{3pt}
\startlongtable
\begin{deluxetable*}{lcccccccc}
\tabletypesize{\scriptsize}
\tablecaption{Derived values for the planets in this study}
\label{tab:sterna_derived}
\tablewidth{\textwidth}
\tablehead{
  \colhead{Planet ID} &
  \colhead{$b$} &
  \colhead{$a (AU)$} &
  \colhead{$R_\oplus$} &
  \colhead{$M_\oplus$} &
  \colhead{$S_\oplus$} &
  \colhead{FPP} &
  \colhead{NFPP}& 
  \colhead{Reliability}
}
\startdata
KOI-4636.01&$1.00\pm0.17$&$0.44\pm0.0096$&$5.01\pm1.44$&$27.2\pm23.3$&$2.01\pm0.128$&$0.391\pm0.067$&$2.93e-01\pm7.53e-02$&$0.61$\\
KOI-463.01&$0.88\pm0.09$&$0.10\pm0.0008$&$1.92\pm0.08$&$5.4\pm2.4$&$2.39\pm0.074$&$0.101\pm0.025$&$0.00e+00\pm0.00e+00$&$--$\\
KOI-87.01&$0.79\pm0.33$&$0.81\pm0.0161$&$1.92\pm0.18$&$5.4\pm2.4$&$0.94\pm0.032$&$0.014\pm0.024$&$1.44e-06\pm1.58e-06$&$0.99$\\
KOI-401.02&$0.81\pm0.29$&$0.58\pm0.0077$&$3.10\pm0.21$&$11.1\pm4.0$&$2.31\pm0.104$&$0.910\pm0.201$&$5.65e-01\pm4.62e-01$&$0.09$\\
KOI-505.05&$0.87\pm0.44$&$0.36\pm0.0060$&$2.78\pm0.23$&$8.9\pm3.5$&$2.33\pm0.068$&$0.028\pm0.013$&$1.40e-05\pm4.58e-05$&$0.97$\\
KOI-1871.01&$0.86\pm0.23$&$0.36\pm0.0065$&$3.03\pm0.18$&$10.6\pm3.7$&$1.65\pm0.056$&$0.003\pm0.007$&$5.93e-07\pm1.51e-06$&$1.00$\\
KOI-5790.01&$0.85\pm0.37$&$0.57\pm0.0095$&$2.15\pm0.18$&$6.0\pm2.3$&$0.85\pm0.034$&$0.607\pm0.107$&$5.92e-01\pm1.09e-01$&$0.37$\\
KOI-2210.02&$0.56\pm0.51$&$0.61\pm0.0119$&$2.90\pm0.25$&$9.8\pm3.9$&$0.58\pm0.043$&$0.021\pm0.008$&$1.00e-07\pm5.30e-08$&$0.98$\\
\enddata
\tablecomments{
* signifies the target star was missing the stellar parameters in the TRICERATOPS Star field which we input manually; ** signifies the planets that could not be processed by TRICERATOPS; -- signifies the planet mass could not be processed by \textit{Spright}, and False alarm reliability $R_{FA}$ is not available.
}
\end{deluxetable*}

\section{appendix}

\newpage
\begin{figure*}[t!]
    \includegraphics[width=1\linewidth]{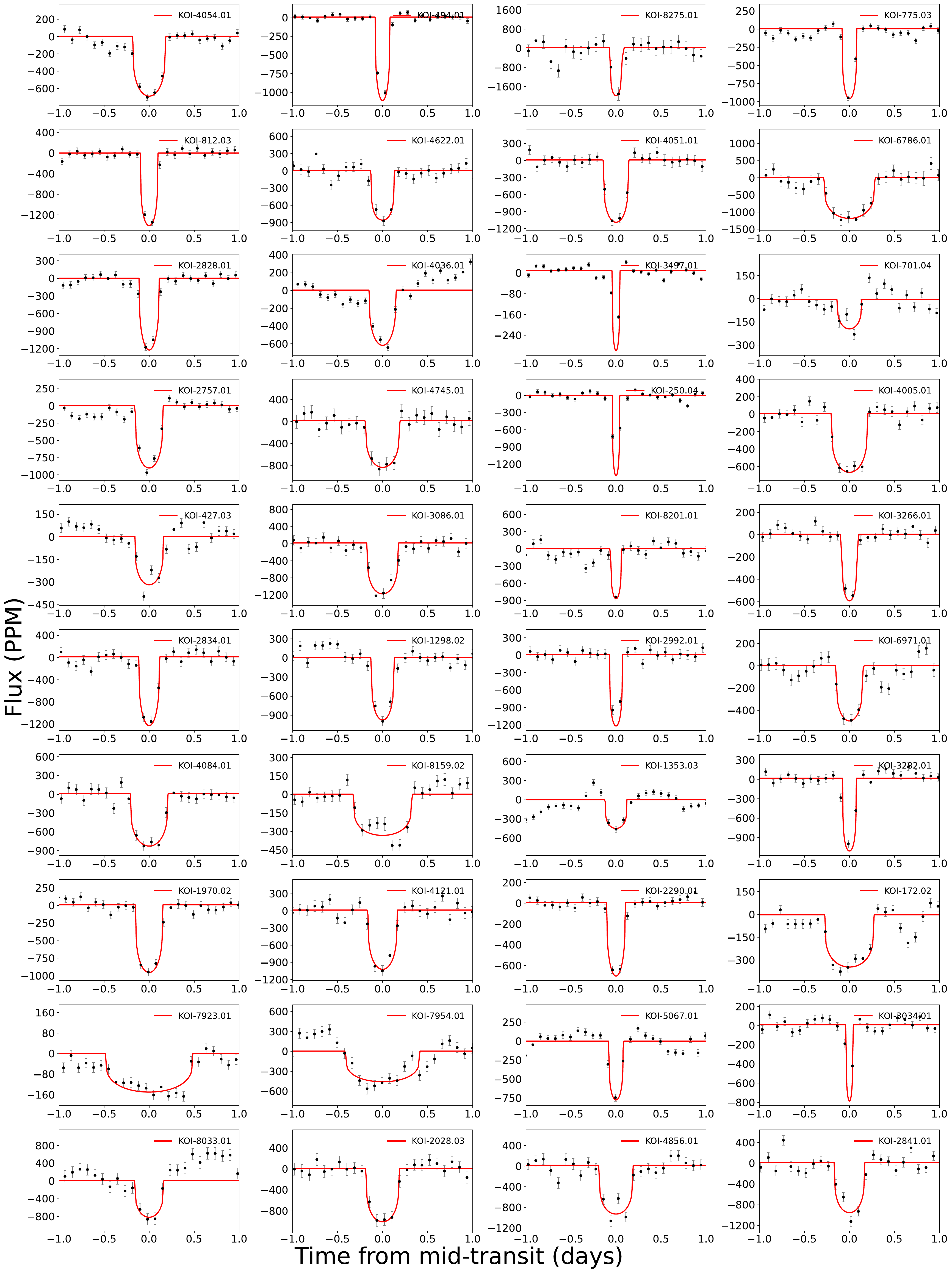}
    \caption{}
    \label{fig:transit_fits1}
\end{figure*}

\newpage
\begin{figure*}[t!]
    \includegraphics[width=1\linewidth]{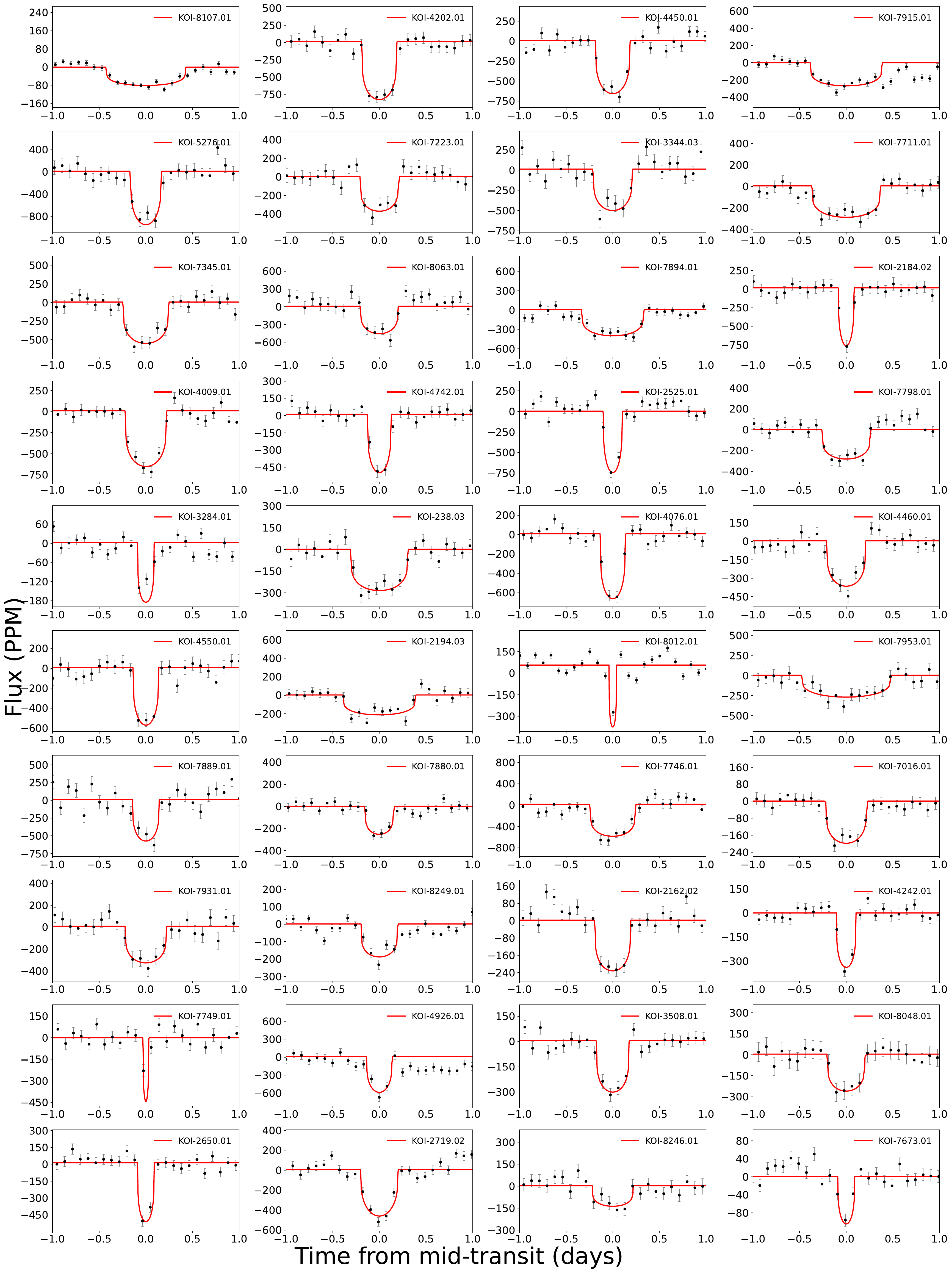}
    \caption{}
    \label{fig:transit_fits2}
\end{figure*}

\newpage
\begin{figure*}[t!]
    \includegraphics[width=1\linewidth]{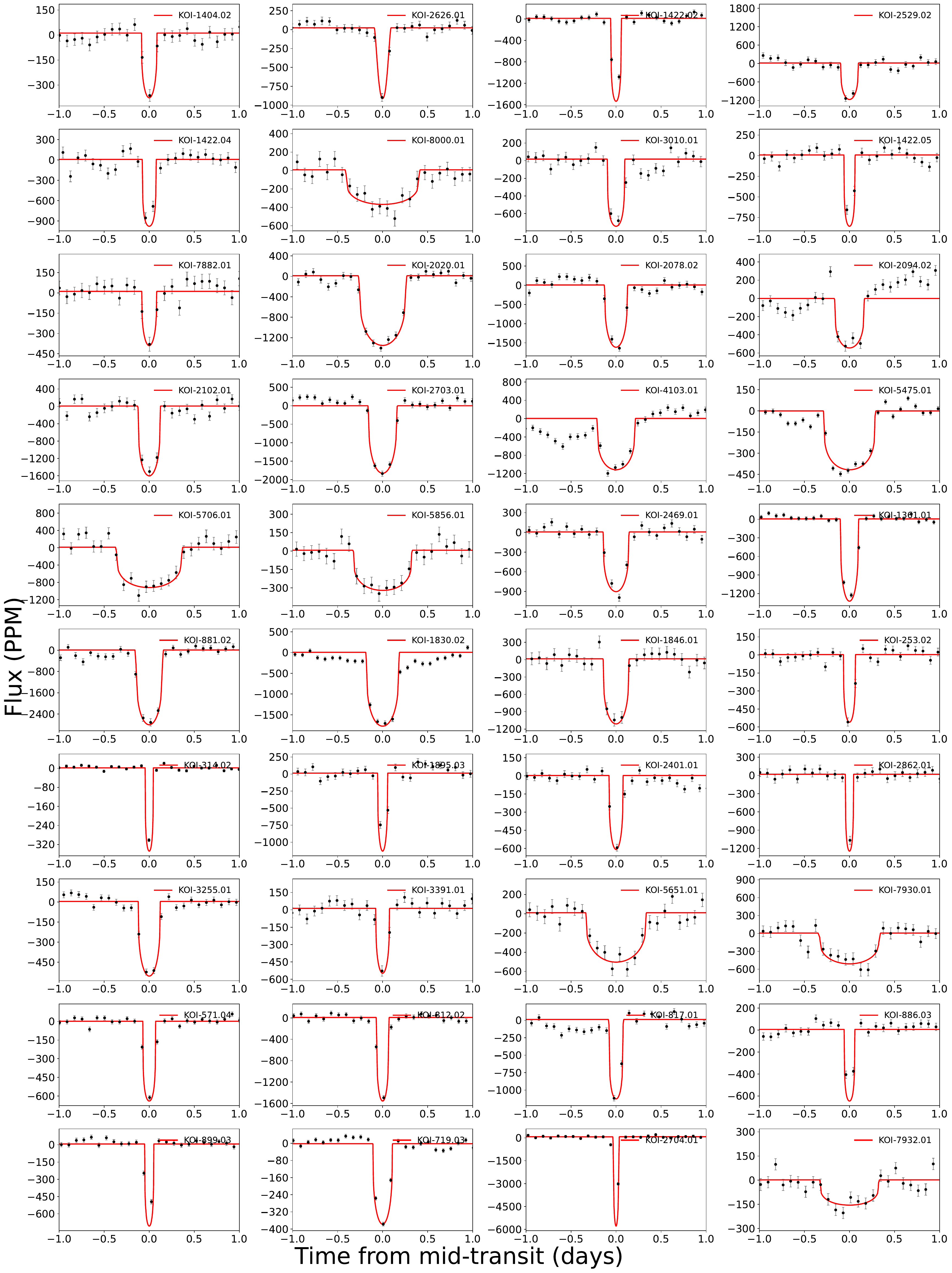}
    \caption{}
    \label{fig:transit_fits3}
\end{figure*}

\newpage
\begin{figure*}[t!]
    \includegraphics[width=1\linewidth]{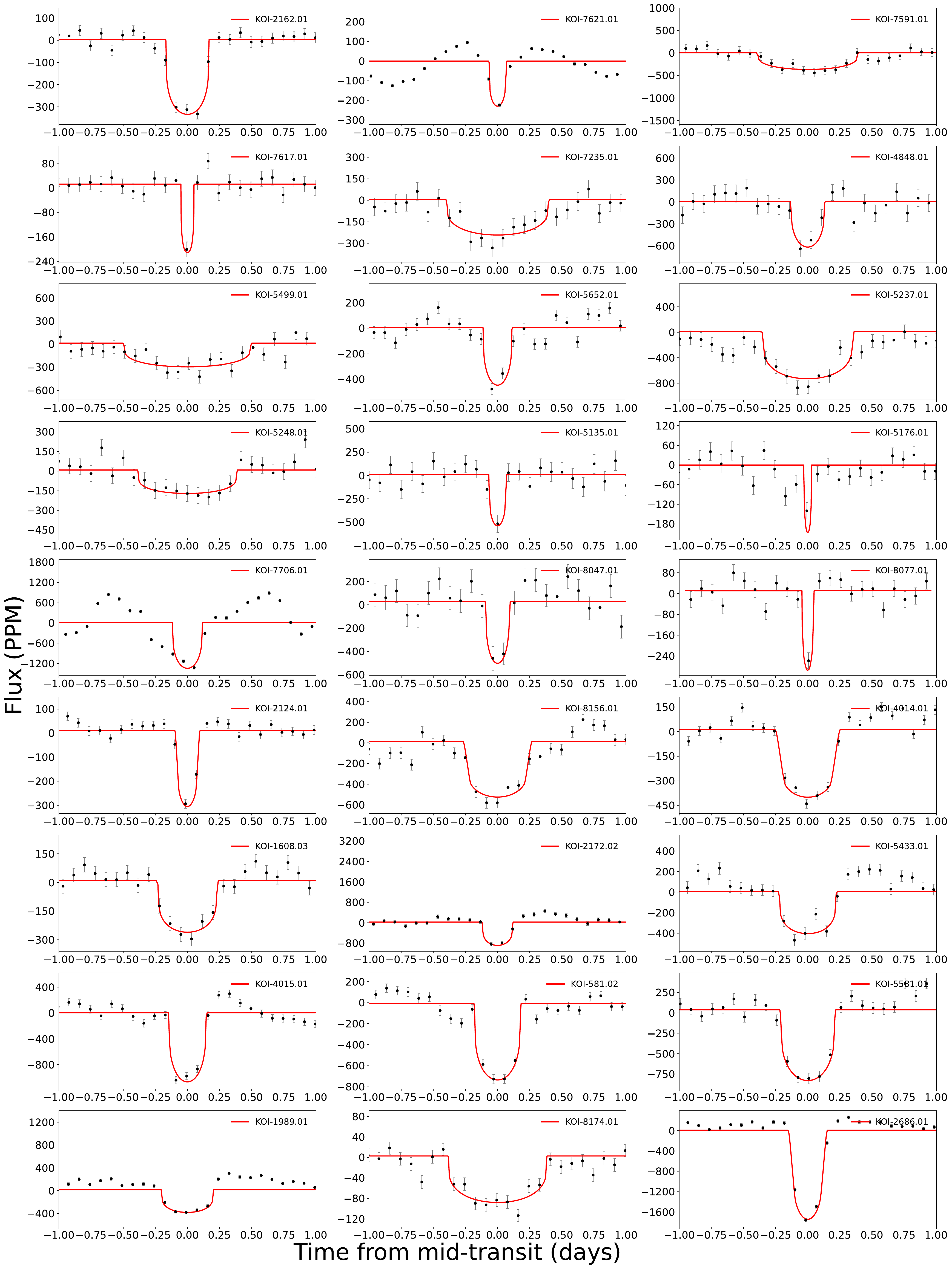}
    \caption{}
    \label{fig:transit_fits4}
\end{figure*}


\clearpage
\bibliographystyle{aasjournalv7}
\bibliography{sample701}

\end{document}